\documentclass[11pt]{article}
\usepackage{amssymb,amsmath,bm}
\usepackage[colorlinks=true, urlcolor=blue,linkcolor=blue, citecolor=blue]{hyperref}
\usepackage{mathrsfs,verbatim}
\usepackage{caption,float}
\usepackage{constants,color}
\usepackage{enumerate}
\usepackage{bbm}
\usepackage{appendix}
\usepackage{graphicx, caption,subcaption}
\begin{document}
\newcommand{\bea}{\begin{eqnarray}}
\newcommand{\ena}{\end{eqnarray}}
\newcommand{\beas}{\begin{eqnarray*}}
\newcommand{\enas}{\end{eqnarray*}}
\newcommand{\beq}{\begin{equation}}
\newcommand{\enq}{\end{equation}}
\def\qed{\hfill \mbox{\rule{0.5em}{0.5em}}}
\newcommand{\bbox}{\hfill $\Box$}
\newcommand{\ignore}[1]{}
\newcommand{\ignorex}[1]{#1}
\newcommand{\wtilde}[1]{\widetilde{#1}}
\newcommand{\mq}[1]{\mbox{#1}\quad}
\newcommand{\bs}[1]{\boldsymbol{#1}}
\newcommand{\qmq}[1]{\quad\mbox{#1}\quad}
\newcommand{\qm}[1]{\quad\mbox{#1}}
\newcommand{\nn}{\nonumber}
\newcommand{\Bvert}{\left\vert\vphantom{\frac{1}{1}}\right.}
\newcommand{\To}{\rightarrow}
\newcommand{\supp}{\mbox{supp}}
\newcommand{\law}{{\cal L}}
\newcommand{\Z}{\mathbb{Z}}

\newcommand{\ucolor}[1]{\textcolor{blue}{#1}}  
\newcommand{\ucomm}[1]{\marginpar{\tiny\ucolor{#1}}}  

\newcommand{\icolor}[1]{\textcolor{green}{#1}}  
\newcommand{\icomm}[1]{\marginpar{\tiny\icolor{#1}}}  

\newcommand{\ccolor}[1]{\textcolor{red}{#1}}  
\newcommand{\ccomm}[1]{\marginpar{\tiny\ccolor{#1}}}  

\newtheorem{theorem}{Theorem}[section]
\newtheorem{corollary}{Corollary}[section]
\newtheorem{conjecture}{Conjecture}[section]
\newtheorem{proposition}{Proposition}[section]
\newtheorem{lemma}{Lemma}[section]
\newtheorem{definition}{Definition}[section]
\newtheorem{example}{Example}[section]
\newtheorem{remark}{Remark}[section]
\newtheorem{case}{Case}[section]
\newtheorem{condition}{Condition}[section]
\newcommand{\pf}{\noindent {\it Proof:} }
\newcommand{\proof}{\noindent {\it Proof:} }
\frenchspacing


%

\title{\bf An overview of time series point and interval forecasting based on similarity of trajectories, with an experimental study on traffic flow forecasting}
\author{İlker Arslan\footnote{MEF University, Mechanical Engineering Dept., Istanbul, Turkey. email: arslanil@mef.edu.tr } \hspace{0.2in} Can Hakan Dağıdır\footnote{Bo\u{g}azi\c{c}i University, Mathematics Dept., Istanbul, Turkey. email: can.dagidir1@boun.edu.tr} \hspace{0.2in}
 \"{U}m\.{i}t I\c{s}lak\footnote{Bo\u{g}azi\c{c}i University, Mathematics Dept.,Istanbul, Turkey. email: umit.islak1@boun.edu.tr}  \hspace{0.2in} } \vspace{0.25in}
\maketitle
\begin{abstract}

The purpose of this paper is to give an overview of the time series forecasting problem based on similarity of trajectories. Various methodologies are  introduced and studied, and detailed discussions on hyperparameter optimization, outlier handling and   distance measures are provided.  The suggested new  approaches   involve variations in both the selection of similar trajectories and  assembling the candidate forecasts.  After forming a general framework, an  experimental study is conducted to  compare the methods that use similar trajectories along with  some other standard models (such as ARIMA and Random Forest) from the literature. Lastly, the forecasting setting is extended to interval forecasts, and the prediction intervals resulting from the similar trajectories approach are compared with the existing models from the literature, such as historical simulation and quantile regression. Throughout the paper, the experimentations and comparisons are conducted via the time series of  traffic flow  from the California PEMS dataset.


\end{abstract}


\section{Introduction}\label{sec:intro}
A time series is a list of data points that are  indexed by  time. Examples of times series include the stock price of a given company, hourly electricity demand, retail sales of a product and the daily temperature at a given city. The problems related to  time series forecasting has become of extreme importance in recent decades due to their wide applications.  We refer to the texts  \cite{hamilton} and \cite{hyndman2}  for two general references on time series  analysis. The purpose of this manuscript is to give an overview of a certain  time series forecasting methodology in detail, namely, the use of similar trajectories.  
 
The experiments below will be  based on the well studied problem of   traffic flow forecasting. The data set we use is the traffic flow data of California PEMS data, and the relevant details are explained  in Section \ref{sec:pdanddataset}. 
Our forecasting approach follows the lines of 
\cite{habt} in which they use similar trajectories to obtain a point forecast for the next time step. The  method in cited work can be summarized as follows. In order to provide a point (or, interval) forecast for time $T  + 1$ when  you are at time $T \in \mathbb{N}$, first look for (windows of) trajectories in the past that are similar to the  recent observations up to time $T$. Then after obtaining some candidates from these similar trajectories, ensemble them via an  appropriate function to reach at a final forecast. Since the process requires the selection and assembly  of nearest neighbors, the method is sometimes known to be the $K$-NN, $K$ referring to the number of similar trajectories that are used in obtaining the forecast. Focusing on the traffic flow case, note that one has seasonal patterns such as daily, weekly and monthly seasonality,  and these need to be taken into account in selection of the nearest neighbors, and relevant discussions will be included below. 

We refer to the papers \cite{habt}, \cite{wang}, \cite{yu} for the use  of similarity of trajectories  in traffic flow.  Among these, as noted earlier, \cite{habt} should be emphasized in our case since our setup will primarily  use the same reasoning behind it.  Of course the idea here is quite natural and has emerged in various other fields other than the traffic flow. For a general discussion of the topic centered around economics, but also applicable to other fields, see  \cite{gilboaetal2}. The papers  \cite{gilboaetal1}  and  \cite{gilboaetal3}  are again on similarity based approaches for predictions, with more technical notes included.    Trajectory similarity is used in various other fields as well, see  \cite{wang2} for prediction of remaining life time, \cite{liu} for location prediction and \cite{zhao}  for predicting daily tourist arrival as examples. The literature  is too  vast to list here, but let us note that most of the work in terms of time forecasting is within the   title $K$-NN.


Although our main interest here is not  the particular case of traffic flow forecasting, we  mention some pointers to the literature in this area too for the sake of completeness.   One classical     approach   in traffic flow forecasting   is the use of  econometric models, such as ARIMA and its variations. See, for example \cite{chen},  \cite{kumar} and \cite{lippi}, for relevant  work. Both machine learning models and deep learning models are extensively used   lately. Let us point out the papers \cite{lippi} on SVR, \cite{kumar2} on ANN and Gradient Boosting on \cite{xia}, as some exemplary work.   Also note that these can be combined with ARIMA and related ones in an additive decomposition manner (\cite{manchun}, \cite{zeng}). Regarding the deep learning models, we   mention a few  work based on  LSTM: \cite{kang}, \cite{tian}, and \cite{yang}.  Lastly note that  \cite{bolsh}, \cite{lippi}  and \cite{miglani} provide comparisons on different aspects of traffic flow forecasting. 

Besides the point forecast problem in time series, we will also be interested in interval forecasts below.  Although these intervals are not studied in the traffic flow case,  they are widely used in several fields in order to supplement the point forecasts.   As an example, electricity price forecasting is a  popular topic in which prediction intervals are widely used, and obtaining such intervals in this case  is  quite challenging due to the spiky behavior of the corresponding time series. See \cite{NW:2015}  and \cite{NW:2018} for relevant discussions. There are various methods for obtaining prediction intervals in the literature, for example,  the historical method, bootstrapping and  distributional prediction intervals  \cite{NW:2018}.    To the best of our knowledge, the attempts using a direct use of similar trajectories in this manuscript are new, and we were not able to find   such references neither in the field of traffic flow  nor in  other  ones.  
  
Overall, our experiments showed that the point forecasting methods   based on similarity of trajectories  are competitive with  various   models used in the literature. We will provide the comparison results between  AR, ARIMA, Random Forest  and a basic version of the similarity method in Section \ref{sec:bchms}. Our experimentations actually included various other models such as gradient boosting, ANN and SVR, but we will not be include a   comprehensive list, since   our main purpose is to experiment,  discuss and compare the methodologies within the  similarity of trajectories framework. As will be clear from the upcoming sections, the similarity approach has  an extensive flexibility in it, and the theoretical classification of problems that can be handled with a certain given method can be interesting. Due to the presence of several options in design of the algorithms here, the computational cost may or may  not be satisfying. The proposed methodologies in general are grounded in the premise that historical patterns sharing similar characteristics can provide valuable insights for estimating future trajectories. Similarity models benefit from the observations from both distant history and outlier (or minority) segments of historical data. With this ability, similarity models can utilize the diversity in the historical data without sacrificing the overall accuracy. In addition to the accuracy and robustness of our forecasts, analyzing similarity models may enhance our understanding of the underlying system dynamics.

The contributions of the current work can be summarized as follows: 
\begin{enumerate}
\item The basics of point and interval  forecasting based on similarity of trajectories are explained in a simple setup in detail in Section \ref{sec:ell2case}.  We hope that this exemplary case gives a clear description of the method  to  a general audience. 
\item Various point forecast mechanisms based on similarity of trajectories are discussed. In particular,   the multi-hourly model and the local regression, that are to be described at length,   approaches provide  the best results among our experiments. The designs are flexible by nature and  seem to be open to further improvement. Also the  forecasts obtained via the direct use of  similar trajectories help in terms of the interpretability. 
\item Results corresponding  to various  distances are collected and listed together, and are compared with the methodologies suggested in the current manuscript.  Some models outside of our approach are also included to make the position of the present attempts clear.  
\item Experimentations for multi-step forecasts via the use of similarity are done, and the results seem to be promising, especially under the setting of seasonal similarity. 
\item Similarity of trajectories approach is used to obtain prediction intervals in time series framework, complementing the point forecasts. Comparisons for this case with other standard methods are  included  as well.  A new algorithm for obtaining  prediction intervals is introduced which  is a mixture of the  so-called historical simulation and the similarity of trajectories methods.
\end{enumerate}

The rest of the paper is organized as follows. The next section is on a description of the traffic flow data set we use, and it further contains the relevant background on the distances of interest,  the forecast evaluation metrics, and sample quantiles. In Section \ref{sec:ell2case}, we stop for a case analysis based on the weighted $\ell^2$ distance.  Section \ref{sec:processing} is devoted to discussions on selection and handling of the candidate trajectories that are to be used for forecasting purposes.  The  point forecast methods   handled in this paper are listed in Section \ref{sec:pointforecasts}. The results of our experiments for point forecasts are also given in this part.   Section \ref{sec:predictionints} includes  the approaches for interval forecasts based on similarity, and also provides the experimentation results for the prediction intervals.    We conclude the paper with a short discussion in Section \ref{sec:conclusion}.
\section{Background}\label{sec:background}

\subsection{Problem Description and Data set} \label{sec:pdanddataset}

Given the first $t$ instances $X_1, X_2, \dots, X_{t-1}, X_t$  of a  time series $X$, the general problem we are interested in is to provide a forecast for $X_{t+1}$. Using similarity of trajectories, this problem can be approached as follows. First, the time series data is transformed into a trajectory-based representation. (Trajectories can be thought as finite  subsequences corresponding to consecutive indices  of sequence of time series. The trajectories to be considered here will be of the same length.) Secondly, similar trajectories are identified using $K$-NN. Then, the succeeding observations of similar trajectories, which are to be called \textit{candidates}, are used for forecasting.

In our experiments, we use real world traffic data from California Freeway Performance Measure Systems (PeMS). In particular, the experiments are done with the flow data for Station 312757. 
The data contains five-minute intervals of  observations such as   total traffic flow and average speed.  The  use  of 15-minute long intervals is advocated in the literature  since using 15 minutes instead of 5  smoothens out the noise while  keeping the data meaningful \cite{lippi}. In our case, we achieve this by combining 3-tuples of non-overlapping five-minute intervals by taking their sum.

\begin{figure}[H]
	\centering
		\includegraphics[width=.99\textwidth, keepaspectratio]{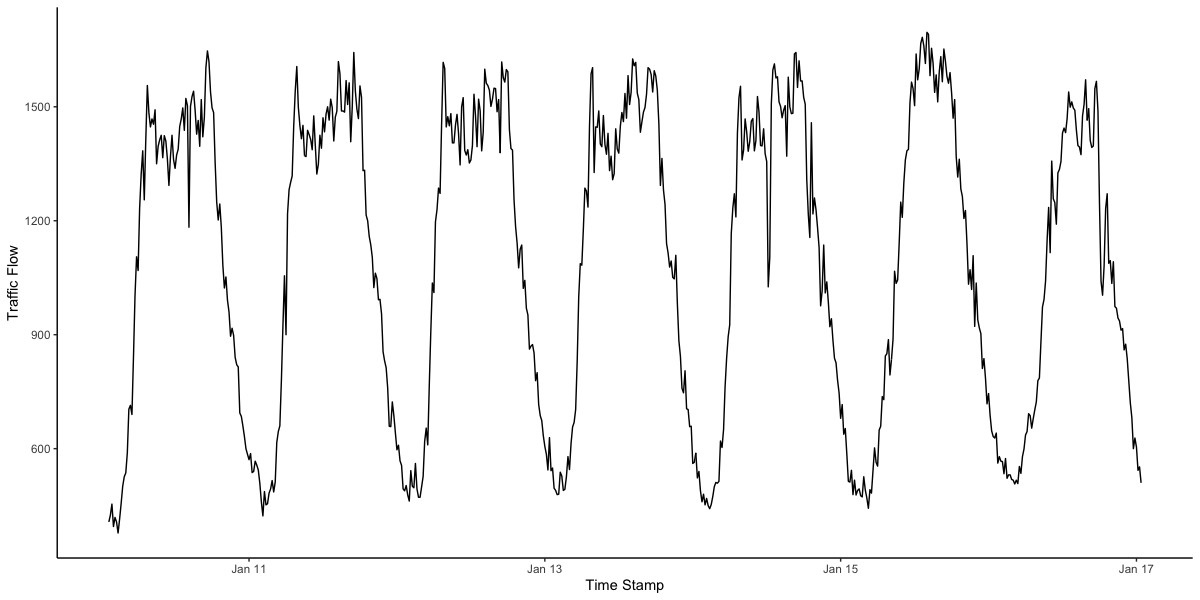}
		\caption{Weekly traffic flow data}
		\label{fig:weeklyts}
\end{figure}


During this study we use the data of  20 months duration from October 2020 to June 2022. 
The data split is designed to contain at least a year long observations in each split to capture seasonality.

We split the data into four parts: tune reference, tune query, test reference and test query. Tune splits are used for optimizing the hyperparameters and test splits are used for measuring the performance. Tune split, containing tune reference and tune query, ranges from October 2020 to February 2022 and test split ranges from February 2021 to June 2022. Due to the nature of similarity based models, query and reference splits are introduced. Query split contains the observations that are desired to be estimated. For each observation in the query split, a reference set, consisting of past observations, is defined. Reference splits contain past observations ranging along at least 12 and at most 16 months long time interval. Tune query split ranges from October 5, 2021 to February 5, 2022 and for each observation in the query split, a reference split is defined to range from October 5, 2020 to the date (and time) of that observation. Test query split ranges from February 2022 to June 2022 with reference split starting from February 2021, again to the date (and time) of that observation. Here we emphasize that the size of reference sets differ for each observation. To be more concrete, for two observations at November 10, 2021, 10:30 AM and January 3, 2022, 6:00 AM, two different reference sets are defined. The former reference set ranges from October 5, 2020 to November 10, 2021, 10:30 AM and the latter reference set again starts from October 5, 2020 but ends in January 3, 2022, 6:00 AM.

The following figure demonstrates the data split done in our study. Note  that the significant change in traffic flow around October 2020 is due to the relaxation of Covid-19 measures.  
\begin{figure}[H]
	\centering
		\includegraphics[width=.9\textwidth, keepaspectratio]{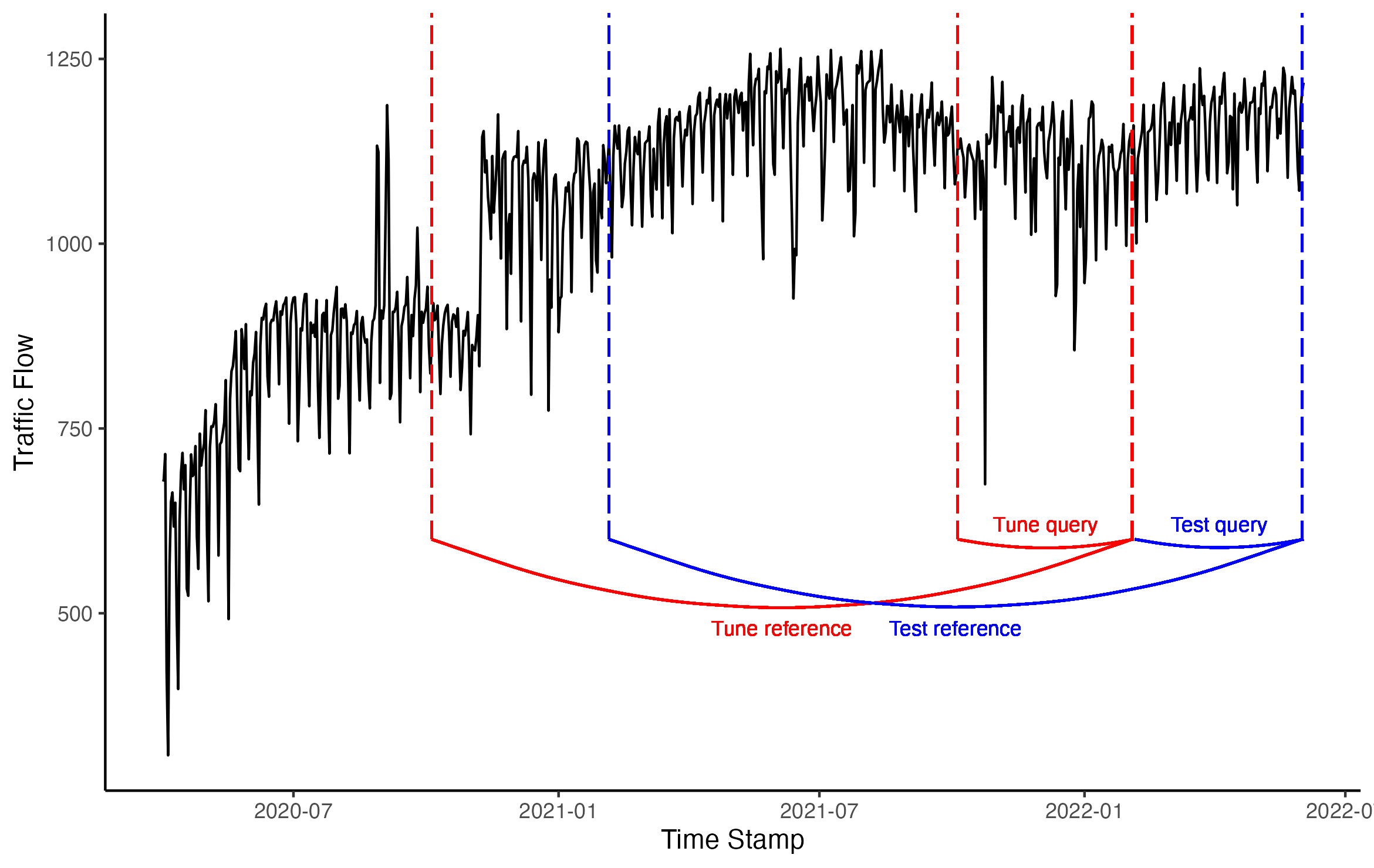}
		\caption{Data split}
		\label{fig:datasplit}
\end{figure}

The missing data in our study is handled in  two different ways. First,  when the length of consecutive missing data is less than one hour, the corresponding flow data from previous week is substituted. Secondly, for longer consecutive periods (which actually only occurs in February 2022) we use the average of the same days and hours of previous three weeks. 



\subsection{Distances}\label{sec:distances}
  
In this subsection we define certain distance functions that can be used for measuring the distance between two trajectories. Before moving further let us provide some references that contain more details about the distances discussed here, and several other ones. \cite{clustersurv} compares various distances, and discuss their use in the clustering framework. \cite{taoetal:2021} gives an overview of distances such as Frechet distance, the dynamic time warping, longest common subsequences. The paper contains four experimental studies for comparison purposes. \cite{suetal:2020} and \cite{DT:2015}   are two other references that discuss various aspects of measures for similarity of trajectories.

Now, let us provide the definitions of  some common distances between trajectories, and discuss them briefly.  Below $\mathbf{x} = (x_1,\ldots,x_L)$ and $\mathbf{y} = (y_1,\ldots,y_L)$ are considered to be $L \in \mathbb{N}$  instances of a real valued time series. 

\textbf{1. $\ell^p$ distance and related.}   For $p \geq 1$,  $p$\textit{-distance} between $\mathbf{x}$ and $\mathbf{y}$ is defined to be $$d_p (\mathbf{x}, \mathbf{y}) = \left(\sum_{i=1}^L |x_i - y_i|^p \right)^{1/p}.$$   $\|\mathbf{x} - \mathbf{y}  \|_p$ is sometimes written  for $d_p (\mathbf{x}, \mathbf{y})$. In this notation the norm of $\mathbf{x}$ is $\|\mathbf{x} \|_p$.

Two special cases should be noted. When  $p = 2$, $\ell^p$ distance is known as the \textbf{Euclidean distance} between $\mathbf{x}$ and $\mathbf{y}$, perhaps the most well-known distance in the literature.  Some authors also consider the \textbf{squared Euclidean distance} $$d_2^2 = \sum_{i=1}^L |x_i - y_i|^2$$ in order to avoid dealing with the  square roots, but that will not be an issue for us. 




The second special case of the $\ell^p $ distance we shall mention  is the \textbf{Manhattan distance} corresponding to $p = 1$: $$d_M (\mathbf{x} , \mathbf{y}) = \sum_{i=1 }^L |x_i - y_i |.$$ Manhattan distance is  also known as $\ell^1$ distance and the taxicab distance.  Computations involving Manhattan distance are faster than the Euclidean distance due to absence of taking square roots. 
  
We lastly note the weighted version of the Euclidean distance. Given weights $w_i \in (0,1)$ adding up to 1, the corresponding \textbf{weighted  Euclideean distance} is defined by 
$$d_{w} (\mathbf{x}, \mathbf{y}) = \left(\sum_{i=1}^L w_i |x_i - y_i|^2 \right)^{1/2}.$$
Our preliminary analysis of the similarity based forecasting methods in Section \ref{sec:ell2case} will be using the weighted Euclidean distance, where the weights will be selected in a linear way according to the recentness of observations \cite{habt}. In present notations, considering $\mathbf{x}$ and $\mathbf{y}$ as the compared trajectories, these weights can be written as $w_i = \frac{i}{L(L+ 1) / 2}$.  
  
\textbf{2. $\sup$ distance}   is defined by $$d_{\infty}(\mathbf{x}, \mathbf{y}) = \max_{i = 1,\ldots, L} |x_i - y_i|.$$ Other names for the $\sup$ distance include the maximum distance and Chebyshev distance. Since the $\sup $ distance looks for uniformly similar patterns, it can be difficult to find similar trajectories in our setting below.

In case one is interested in only the similarity of certain coordinates   of $\mathbf{x}$, $\mathbf{y}$, one may consider the  maximum in previous definition over a  subset  $S $  of $\{1,\ldots, L\}$. Below we will focus on a specific case which is to be called  the head-tail distance.  Letting $\ell_1, \ell_2 $ be two positive integers for which $\ell_1 + \ell_2 \leq L - 1$,   the ($\ell_1$-$\ell_2$) \textbf{head-tail distance}, based on some underlying distance $d$ (such as Euclidean), between the given two windows is defined by 
$$d_{HT}(\mathbf{x}, \mathbf{y}) = \sum_{i=1}^{\ell_1} d(x_i, y_i)   +  \sum_{i=L-\ell_2 + 1}^{L} d(x_i, y_i).$$
Note that one may further  prefer to choose the distances in these two summations differently.

\textbf{3. Cosine distance}    is defined to be $$d_{\cos} (\mathbf{x}, \mathbf{y}) =  1 - \frac{\sum_{i=1}^L x_i y_i}{\|\mathbf{x} \|_2 \|\mathbf{y} \|_2 }.   $$   Writing $\theta$ for the angle between $\mathbf{x}$ and $\mathbf{y}$, $d_{\cos} (\mathbf{x}, \mathbf{y}) = 1 - \cos \theta$.
 The cosine distance is known to work effectively in higher dimensions, and it is used in areas such as natural language processing and   text mining. 
 

A closely related distance is  \textbf{Pearson correlation} defined by $$d_{\mathrm{Pearson}} (\mathbf{x}, \mathbf{y}) = 1-  \frac{\sum_{i=1}^L (x_i - \overline{x}) (y_i - \overline{y})}{ \left( \sum_{i=1}^L (x_i-\overline{x})^2 \right)^{1/2}  \left( \sum_{i=1}^L (y_i-\overline{y})^2 \right)^{1/2}}$$
Note that correlation distance is not a metric. When $\overline{x} = \overline{y} = 0$, $\|x\|_2 = \|y\|_2 = 1$, $d_{\mathrm{Pearson}} (\mathbf{x}, \mathbf{y})$ simplifies to $1 - \sum_{i=1}^L x_i y_i$. Then under the same assumptions, we have $d_{\mathrm{Pearson}} (\mathbf{x}, \mathbf{y})  = d_{\cos} (\mathbf{x}, \mathbf{y})$, and also  
$$
d_2^2 (\mathbf{x}, \mathbf{y} ) = \sum_{i= 1}^L (x_i - y_i)^2 =  \sum_{i= 1}^L x_i^2 + \sum_{i= 1}^L  y_i^2 - 2 \sum_{i= 1}^L  x_i y_i = 2  d_{\mathrm{Pearson}}(\mathbf{x}, \mathbf{y}).$$
Let us note that another distance similar to Pearson correlation can be defined by using   Spearman correlation. The main difference from Pearson correlation is that the variables of interest in this case are rank-ordered. 

\textbf{4. Canberra distance} is defined to be $$d_{\mathrm{Canberra}} (\mathbf{x}, \mathbf{y}) = \sum_{i = 1}^L \frac{|x_i - y_i|}{|x_i|+|y_i|}.$$ This distance is a weighted version of the $\ell^1$
 distance, where the weights help robustness to outliers. It is used in various areas including clustering and classification, and has several variations which we will not discuss here. 

\textbf{5. Others.} There is vast list of  other similarity measures in the literature that can be adapted to measure similarity of trajectories. These include but are not limited to mean character difference, Frechet distance,  coefficient of divergence, matching subsequence counts, dynamic time warping, longest common subsequence (\cite{bc:1994}, \cite{io}, \cite{clustersurv},   \cite{sankoff}, \cite{taoetal:2021}).  In our study we had some preliminary   experiments  with the longest common subsequences approach.
The length of the longest common subsequences is a classical measure of similarity for discrete sequences.  It is used, for example, in molecular biology for DNA comparison,   computer science for binary sequence comparison, among others. See \cite{capocelli},  \cite{pevzner} and \cite{sankoff} for various applications of the longest common subsequences as well as algorithmic discussions. Recent theoretical advances on the common subsequences problem can be traced in   \cite{islakhoudre}. The theoretical background can also be carried over to a score function setting which can be used to treat continuous data. See \cite{GHI} for relevant discussions and references.

Following the discussion in \cite{taoetal:2021}, the variant of the longest common subsequences for comparison of continuous trajectories can be  defined  as a dynamic programming algorithm as follows. Letting  $\mathbf{x}$ and $\mathbf{y}$ be as before, the length of the longest common subsequence $LC(\mathbf{x}, \mathbf{y})$ is defined by 
$$
LC(\mathbf{x}, \mathbf{y}) =
\begin{cases}
0, & \text{if } \mathbf{x} \text{ or } \mathbf{y} \text{ is empty}, \\
1 + LC(\mathbf{x}[1:k-1], \mathbf{y}[1:m-1]), & \text{if } |x_k - y_m|< \epsilon \text{ and } |k -m | \leq \delta, \\
\max\{ LC(\mathbf{x}[1, k -1], \mathbf{y}), LC(\mathbf{x}, \mathbf{y}[1:m-1]) \}, & \text{otherwise. }
\end{cases}
$$
Here, the values of the $\epsilon$ and $\delta$ are to be chosen by the researcher. The choice of $\delta = L - 1$   yields the natural analogue of the longest common subsequences in a continuous setup. Our basic experiments with the longest common subsequences yielded poor performance, and we therefore do not list them below.  A variant of this subsequence approach is   the dynamic time warping for which relevant discussions   can be found in   \cite{taoetal:2021}  and  \cite{bc:1994}. We hope to get back to the longest common subsequence, dynamic time warping and their variations in more detail in a subsequent work.  

\subsection{Evaluation metrics}\label{sec:evamets}
  In this subsection we briefly describe the evaluation metrics  for point and interval forecasts that are used in this manuscipt. 
  \vspace{0.1in}

\noindent \textbf{Point forecasts.}  Below we will be using the mean absolute error (MAE) and the mean absolute percentage error (MAPE) for the evaluation of our point forecasts. One may refer to \cite{hyndman2} and \cite{hyndman} for    general discussions on   the point forecast evaluation metrics. 
 MAE corresponding to the  forecasts $\hat{y}_1,\ldots,\hat{y}_m$ for the realized sequence $y_1,\ldots,y_m$ is defined by $$MAE = \frac{1}{m} \sum_{i=1}^m |\hat{y}_i - y_i|.$$ 
The mean absolute percentage error (MAPE)  is defined by $$MAPE = \frac{1}{m} \sum_{i=1}^m \left| \frac{ \hat{y}_i - y_i}{y_i} \right|.$$
Some researchers in the traffic literature use other related metrics such as $MAE_{100}$  for the traffic flow prediction, which basically ignores the time instances $t$ for which $y_t < 100$. For example, see \cite{lippi}. We do not consider a similar strategy here since   minimal flow in our case is not too small, and making use of such a metric does not make much difference than using the standard  $MAE$. 

Once point forecasts are obtained via two distinct models, one may also like to know whether one method is significantly  different than (better or worse) the other one. For this purpose we will be using the Diebold-Mariano test, see \cite{DM:1995} and \cite{DM:2015} for details.
  
   \vspace{0.1in}

\noindent  \textbf{Interval forecasts.}  The two criteria used for comparison of performance of prediction interval mechanisms below are (i) Unconditional coverage, and (ii) Winkler score.  Let us now briefly go over these. 

\begin{itemize}
\item \textbf{Unconditional coverage.} The simplest metric one may consider for a prediction interval is the unconditional coverage, which   is just the proportion of samples in test data covered by our prediction intervals. Formally, given the prediction intervals $[\hat{L}_t, \hat{U}_t]$, $t = 1,\ldots,S$, with the corresponding actual values $y_1,\ldots,y_S$, the unconditional coverage is defined to be $$\mathrm{UC} = \frac{1}{S} \sum_{t = 1}^S \mathbf{1} (y_t \in [\hat{L}_t, \hat{U}_t]),$$ where $\mathbf{1}$ is the characteristic function giving $1$ if the input is true, and 0 otherwise.  For $\alpha \in (0,1)$, a $(1-\alpha)100\%$ prediction interval construction will be considered to work \textit{well} if the $\mathrm{UC}$ is \textit{close} to $(1-\alpha)100\%$. 

Note that there is also the conditional coverage of prediction intervals which also penalizes the clustering of 0's and 1's (for the sequence $ \mathbf{1} (y_t \in [\hat{L}_t, \hat{U}_t])$). Such clusterings may occur in presence of data with  significant outliers. The electricity price forecasting  with its spiky behavior is one such example, see \cite{NW:2018}. We will not be reporting on conditional coverage below.  

\item \textbf{Winkler score.} Besides the obvious coverage probabilities, one other natural criteria to consider is  the  length of prediction interval. One naturally wishes to have the prediction intervals as short as possible (keeping the UC around the desired level). For this purpose we will be using  Winkler score and and  penalize the longer prediction intervals \cite{winkler}. Before giving the definition for the Winkler  score of a prediction interval, let us note that the constructions of our prediction intervals are symmetric around the point forecasts. Now, for $\alpha \in (0,1)$, for the time series $y_t$, $t \geq 1$, and for the corresponding prediction intervals $[\hat{L}_t, \hat{U}_t]$, we follow \cite{NW:2018} and  define the corresponding Winkler score by setting 
$$
W_t =
\begin{cases}
\hat{U}_t - \hat{L}_t & \text{if } y_t \in [\hat{L}_t, \hat{U}_t], \\
\hat{U}_t - \hat{L}_t  + \frac{2}{\alpha} (\hat{L}_t - y_t) & \text{if } y_t < \hat{L}_t \\
\hat{U}_t - \hat{L}_t  + \frac{2}{\alpha} (y_t - \hat{U}_t ) & \text{if } y_t > \hat{U}_t. 
\end{cases}
$$
\end{itemize}
As is clear from the definition, Winkler score penalizes the cases where the actual value of the time series is outside the prediction interval, and in this case, it is proportional to the distance between the actual value and the closest point on  the prediction interval.  We may then define the average Winkler score over the time interval $1,\ldots,S$ to be $$\overline{W} = \frac{1}{S} \sum_{t=1}^S W_t.$$ 
It is also clear from these definitions that a lower Winkler score is an indicator of a better prediction interval performance.

%
%
%
%

\subsection{Sample Quantile}\label{subsec:sample_quantile}

For our prediction interval models, we need to estimate upper and lower quantiles from a sample. There  are multiple ways to define the corresponding sample quantiles, and in this section we fix one that is to be used throughout the paper.  When the sample size is small (e.g. less than 50), a  precise calculation of the quantile would be more desirable. Due to this,  the (precise) sample quantile is defined as follows. Assume we have an ordered sample $\{ a_1, a_2, \dots, a_N \}$ of size $N$, such that $a_i \leq a_{i+1}$ for each $i$. Then $i$-th element of the sample $a_i$, corresponds to the $\frac{i}{N+1}$-th quantile. With that reasoning  we define the $q$-th  sample quantile in the following way. For $q \in [0,1]$,
$$
a_{(q)} = \left\{ \begin{array}{ll}
 a_{q(N+1)}, &\mbox{ if $q(N+1) \in \mathbb{Z}^{\geq 0}$} \\
  c_1 a_{\lfloor q(N+1) \rfloor} + c_2 a_{\lceil q(N+1) \rceil}, &\mbox{ otherwise}
       \end{array} \right.
$$
where $c_1 = 1 - (q(N+1) - \lfloor q(N+1) \rfloor)$ and $c_2 = q(N+1) - \lfloor q(N+1) \rfloor$.

\section{Analysis for weighted Euclidean  case}\label{sec:ell2case}

This section contains a detailed  analysis of  the general methodology in this paper in a specific setting. Here, following \cite{habt},  we specialize on the case where  the  similar trajectories are  obtained via a weighted variant of $\ell^2$ distance as explained in Section \ref{sec:distances} (In particular, the weights will be $w_i = i / (L(L+1)/2)$, $i = 1,\ldots,L$).   In later sections, there will be various approaches in use, but they share certain common characteristics. Focusing on a special case  will retain us repeating several issues, and we hope that  the study in this section will clearly expose the ideas which are involved in general. 



Let us start by recalling the general reasoning of the similarity of trajectories approach.
First, we generate trajectories based on lagged observations of traffic flows. Then we identify the nearest neighbors for each trajectory and save the corresponding flow values as candidates. These candidates should closely resemble the flow that we aim to predict, but as usual they may contain outliers, which are to be handled before obtaining    point and  interval forecasts via   the candidate sets. After the outlier removal is done, the remaining candidates are used to arrive at the final forecasts. The following diagram provides a broad description of the methods we use throughout this paper.


\begin{figure}[H]
	\centering
		\includegraphics[width=.99\textwidth, keepaspectratio]{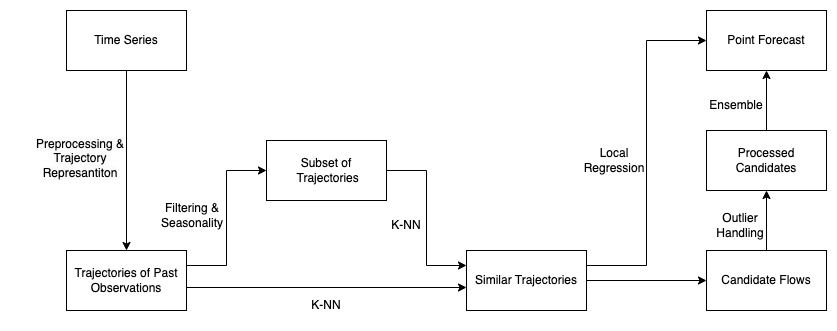}
		\caption{Flowchart showing the general similarity of trajectories methodology}
		\label{fig:flowchart1}
\end{figure}



\subsection{Obtaining the Nearest Neighbors}

The selection of the nearest neighbors is   one of the most critical parts of historical similarity methods. In order to have promising results, it  is crucial that the nearest neighbors have a common and meaningful pattern. Furthermore, the number of nearest neighbors used must be sufficient enough to provide a good view. Keeping these in mind,  we must select optimal values for the three primary hyperparameters: 
\begin{itemize}
\item Window size, $L$, for defining the trajectories, 
\item Distance function, $d$, for measuring similarity, 
\item Number of nearest neighbors, $K$, for obtaining as many neighbors as required in order to have  satisfactory relationships. 
\end{itemize}  
Recall that  we  fixed the distance measure to a weighted Euclidean distance throughout this section. The time windows that are to be considered in this section will be ranging  from 30 minutes to 5 hours (i.e., 2 to 20 past observations) resulting from  15-minute intervals in our data set. We  save up to 200 nearest neighbors for  forecasting purposes. Further hyperparameter selection discussions will be given in the sequel. 

\begin{figure}[H]
	\centering
		\includegraphics[width=.8\textwidth, keepaspectratio]{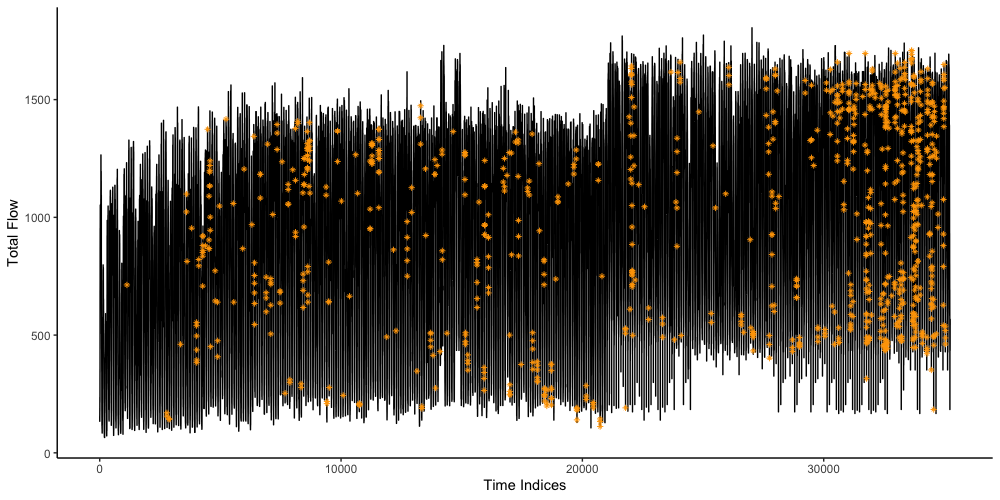}
		\caption{A sample distribution of the nearest neighbors}
		\label{fig:kneighbors}
\end{figure}

Now, before proceeding further, let us include an independent note on  the distribution of  nearest neighbors. One  may   expect to find   seasonal or other distinguishable patterns in historical search, but our experimentations do not  reveal any visible pattern at all. In particular, in Figure \ref{fig:kneighbors}, we present an exemplary distribution of the 10 nearest neighbors for 100 specific consecutive trajectories (To find the nearest neighbors in this specific case, we set the trajectory length to 16 and use the weighted Euclidean distance). Although the most of the nearest neighbors were from more recent observations, we observed that there were several candidates from as far back as a year ago. This may imply that extending the interval of the data may be beneficial for our models.


%
%



\subsection{Benchmarks}\label{sec:bchms}


In this section we divert from the main discussion in order to   describe our benchmark models. We have chosen four methods for this purpose, namely, Naive (or, Random Walk), Autoregressive model (AR), Autoregressive Integrated Moving Average (ARIMA) and Random Forest. 

The naive approach simply uses  the current flow value as the forecast for the next time step ($X_t = X_{t-1}$). This method sets the baseline performance results and any proposed method is expected to outperform it. 
A slightly more complicated version of the stantard naive method is the seasonal naive, in which the forecast we give is based on an averaging with respect to certain previous time instances (such as previous days or weeks),  taking seasonality into account. Since the results of seasonal naive were not close to the ones that are to be described below, we will only provide the ones for standard naive approach. 

 AR and ARIMA models are econometric models that are widely  used in time series analysis and forecasting. AR models use previous values of the target variable to forecast future values. In particular, if the series of interest is $X_t$, then the AR($p$) model can explicitly  be   written as $X_t = \sum_{i=1}^{p} \alpha_i X_{t - i} + \epsilon_t$, where $\epsilon_t$ is white noise and  $\alpha_1,\ldots,\alpha_p$ are the parameters of the model. 
On the other hand,  ARIMA and SARIMA models combine autoregressive and moving average components to account for trends and seasonal patterns in the data. These models are capable of capturing data patterns and dependencies, such as day-of-week effects and hourly variations, and have the potential to provide relatively accurate predictions even with a limited data. As a result, AR and ARIMA serve as robust and simple benchmarks methods for our research. We refer to \cite{hamilton} and \cite{hyndman3}  for further details.

Random Forest (RF) is widely recognized for its effectiveness in capturing nonlinear relationships within high-dimensional data. In short-term traffic flow forecasting literature, Random Forest has demonstrated competitive performance as an ensemble learning method \cite{zhang_rf}. Given the resemblance in methodological approaches between our proposed model and Random Forest, together with the demonstrated efficacy of Random Forest in previous applications, we selected it as the primary benchmark here.


The benchmark results we obtained are summarized in the following table.


\begin{table}[H] \centering 
\begin{tabular}{@{\extracolsep{5pt}} ccccccc} 
\\[-1.8ex]\hline 
\hline \\[-1.8ex] 
Model & MAE & MAPE \\ 
\hline \\[-1.8ex] 
Naive & 54.74 / 56.73 & 5.35 / 5.36 \\
AR & 50.34 / 52.62 & 4.92 / 4.94  \\
ARIMA & 49.34 / 51.83 & 4.82 / 4.83  \\
SARIMA  & 42.09 / 46.08 & 4.18 / 4.39  \\
RF & 43.58 / 46.85 & 4.35 / 4.40  \\
Seasonal RF & 42.91 / \textbf{45.99} & 4.31 / \textbf{4.32}  \\
\hline \\[-1.8ex] 
\end{tabular} 
  \caption{Benchmark results for point forecast}
  \label{tab:point_bench} 
\end{table} 

Note that in  Table \ref{tab:point_bench} and below,  the result on   left-hand-side   shows the tune performance and the one on right-hand-side the test performance. Now we continue with some comments on the results of benchmark models, and on some notes relevant to their implementation. First let us mention the setup for  ARIMA and AR  models which are seen to improve the naive model. For the AR model, we identified the  optimal value of $p$ to be 9,  yielding a mean absolute error (MAE) of $52.62$. The ARIMA model, with optimal values of $p=9$, $d=0$, and $q=3$, exhibited superior performance with an MAE of $51.83$. Finally, the SARIMA model demonstrated the best performance among the three benchmarks, with optimal hyperparameters $p=9, d=0, q=3, P=3, D=0, Q=2$, and seasonal frequency $S=96$, achieving an MAE of $46.08$. The results highlight the importance of seasonality.

Random Forest models consist of several hyperparameters. We refer to \textit{ranger} R package \cite{ranger_package} for naming and detailed description. In our models, the ``mtry" hyperparameter that determines the number of features randomly sampled at each split, ranges from 4 to 11. Meanwhile the ``min.node.size" hyperparameter, determining the minimal node size to split at, varies from 2 to 20. There are various studies regarding the ``num.trees" hyperparameter that determines the number of trees in each forest. These studies imply that higher values simply yield better performance. However, our experiments indicate that the improvement in performance becomes insignificant after the number of trees exceed 150. Therefore, for our experimentation, the maximum value for ``num.trees" in the grid search is set to 200. Also, the models trained with minimizing mean absolute error performed better, compared to the ones that minimize mean square error. \label{par:rf_details}

We had two main approaches for the random forest models. The first one only used lagged observations as the features. We tried various lengths for window size and stopped at 30 since no further improvement was observed in    the accuracy. The first approach yielded an MAE of $47.01$, performing better than non-seasonal autoregressive benchmarks but worse than SARIMA, implying a motivation for a random forest model with seasonal parameters (SRF). In the second approach, we also included observations from one day and one week before. Again we conducted a similar grid search for window sizes and other hyper-parameters. SRF performed best among other benchmarks with MAE of $45.85$.

Lastly, the train/test split of benchmark models differ from the splits of similarity models. We keep the size of test split the same, 4 months, and use 12 months of observations for training.\footnote{As an additional note, using larger sets for training resulted in worse test scores.}

\subsection{Point Forecasts: Similarity of Trajectories}

There are various issues to be considered in providing point forecasts via similarity of trajectories. In this section, focusing on the aforementioned weighted Euclidean distance used in this section,  we give a general overview of these, by focusing on   hyperparameter tuning, outlier handling and multi-step forecasting. 

\vspace{0.1in}

\noindent \textbf{Tuning Hyperparameters.}   Hyperparameter optimization is a crucial step in forecasting problems.   Recall in our case that the hyperparameters  are the distance function $d$, the number of neighbors selected $K$ and the window size $L$. Since the distance function is already fixed, we are left with adjusting $K$ and $L$.  Similar tuning is  required to be done for the  forecasting mechanisms described  in Section \ref{sec:pointforecasts}, but not all the details will be presented there.

In our treatment below in this section, the optimal values of window size $L$ is searched from 2 to 20. Then for each $L$, average of $K$ candidates from nearest  neighbors is used as the forecast,  where $K$ ranges from 5 to 200. The following heat map  summarizes our experiment results for this range of $K$ and $L$. 

\begin{figure}[H]
	\centering
		\includegraphics[width=.45\textwidth, keepaspectratio]{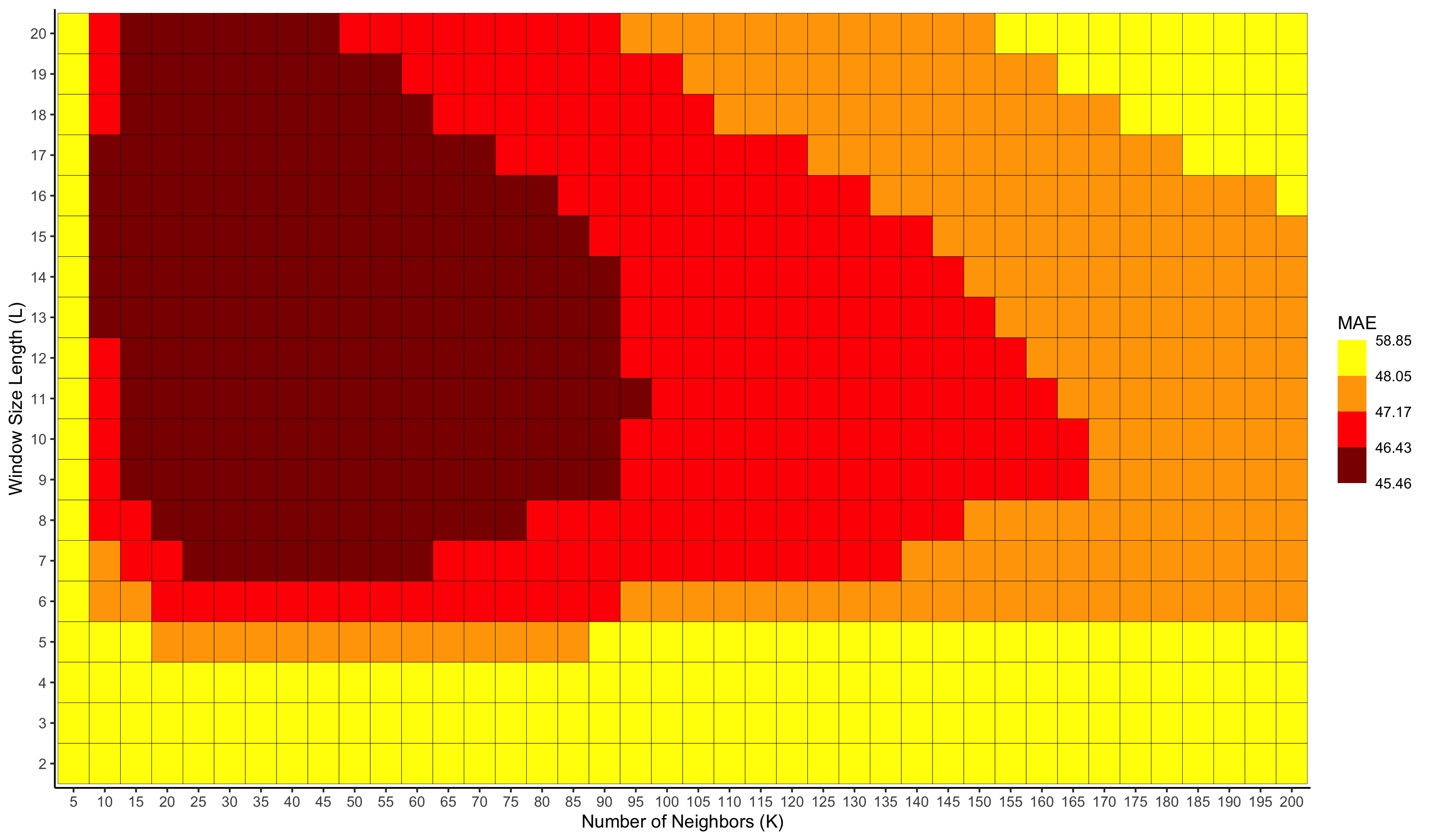}
		\includegraphics[width=.45\textwidth, keepaspectratio]{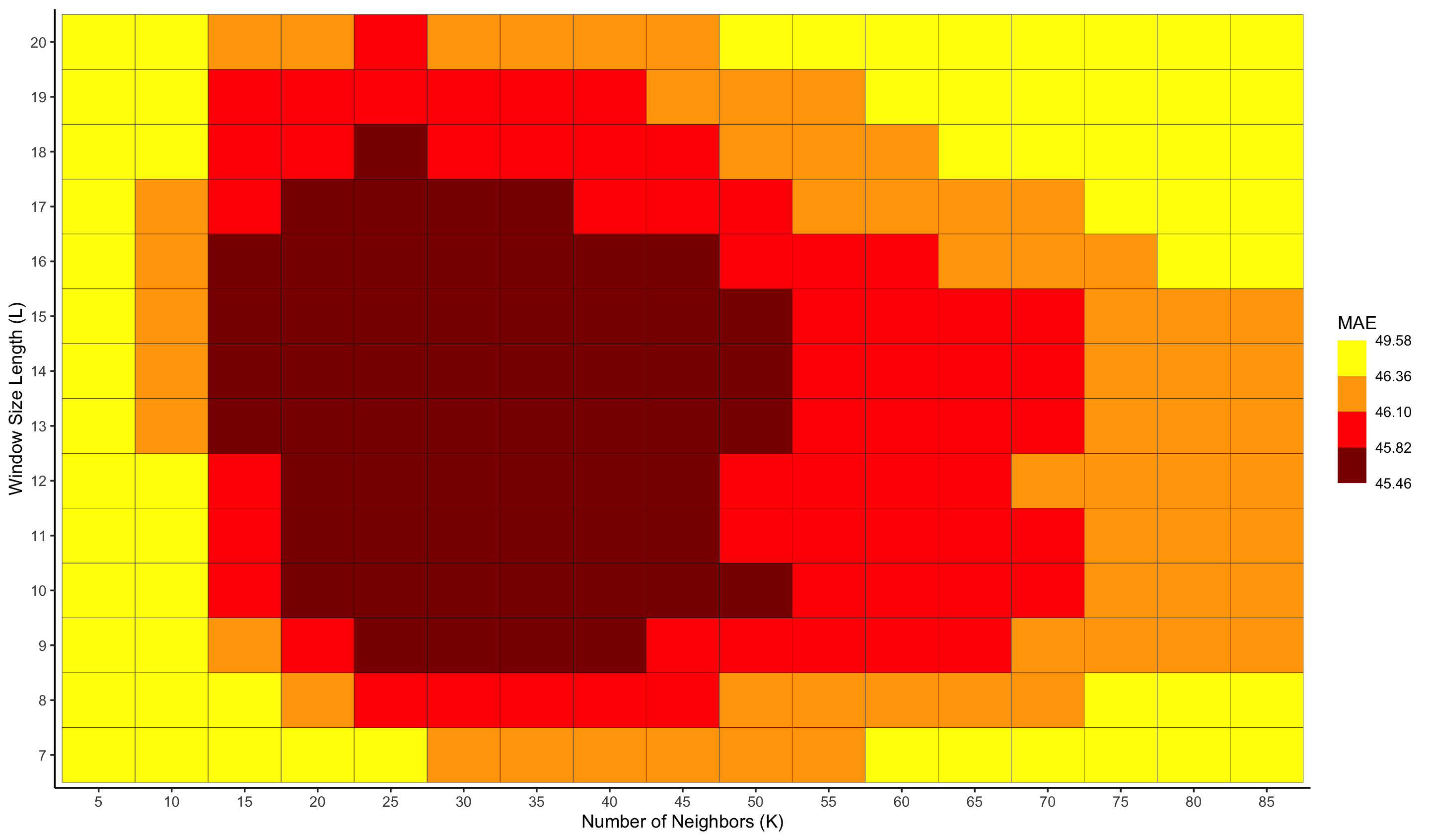}
		\caption{Initial Grid Searches for $K$ and $L$. Left figure shows a wider search. Right figure focuses on the better performing hyperparameters.}
		\label{fig:kl_grid1}
\end{figure}

In Figure \ref{fig:kl_grid1}, the left hand side shows a general view, and then with zooming in on the right hand side we see a nicely shaped region containing the optimal hyperparameters. This region suggests us to make a second grid search with more frequent values for $K$ between 10 and 50, and their pairings with $9 \leq L \leq 17$. Note that we were conducting this initial experimentation on the tuning split, so the results differ from the final results given in Section \ref{sec:pointforecasts}.


\begin{figure}[H]
	\centering
		\includegraphics[width=.48\textwidth, keepaspectratio]{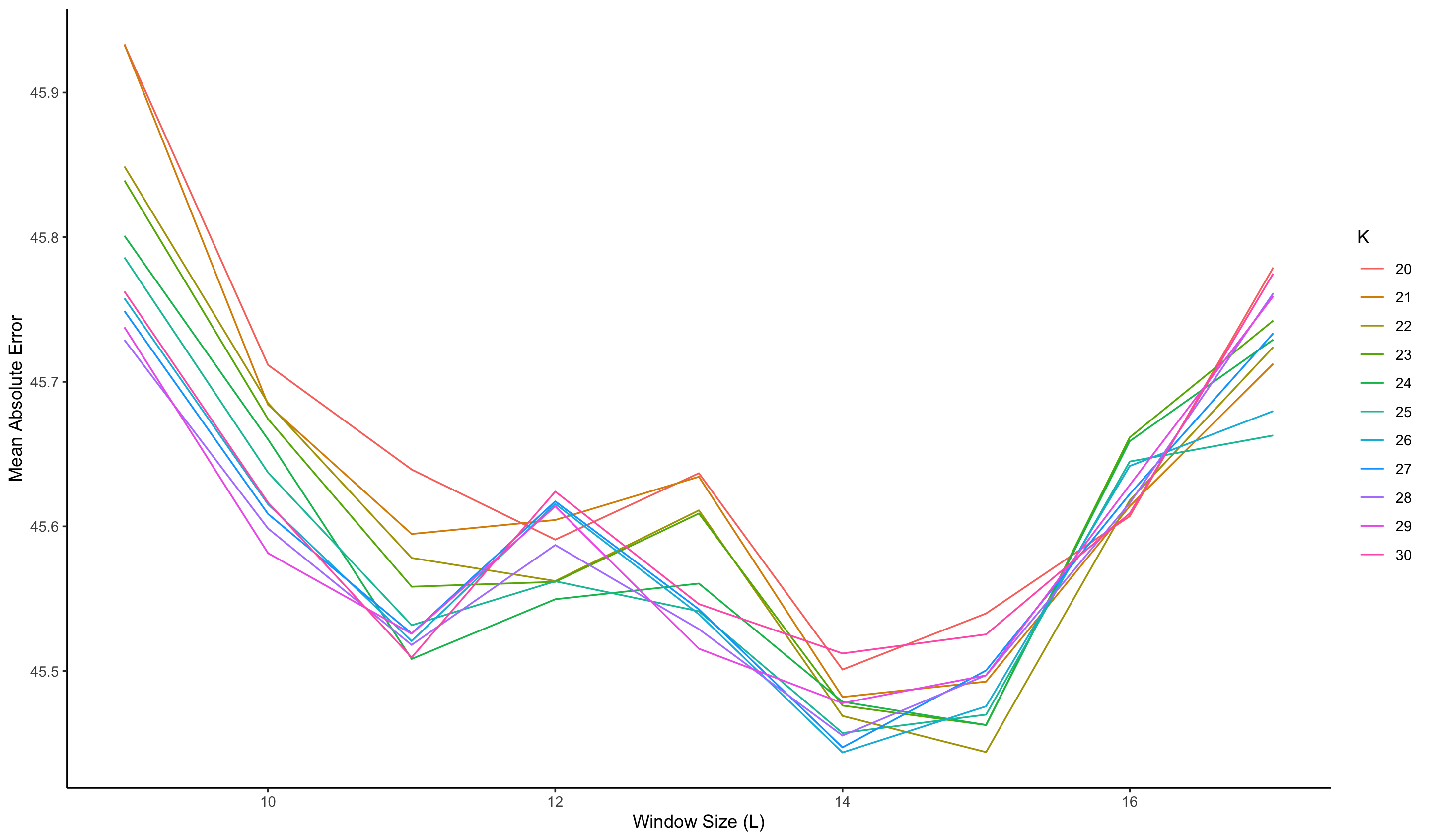}
		\caption{Best $L$ values}
		\label{fig:kl_tick1}
\end{figure}

\begin{figure}[H]
	\centering
		\includegraphics[width=.48\textwidth, keepaspectratio]{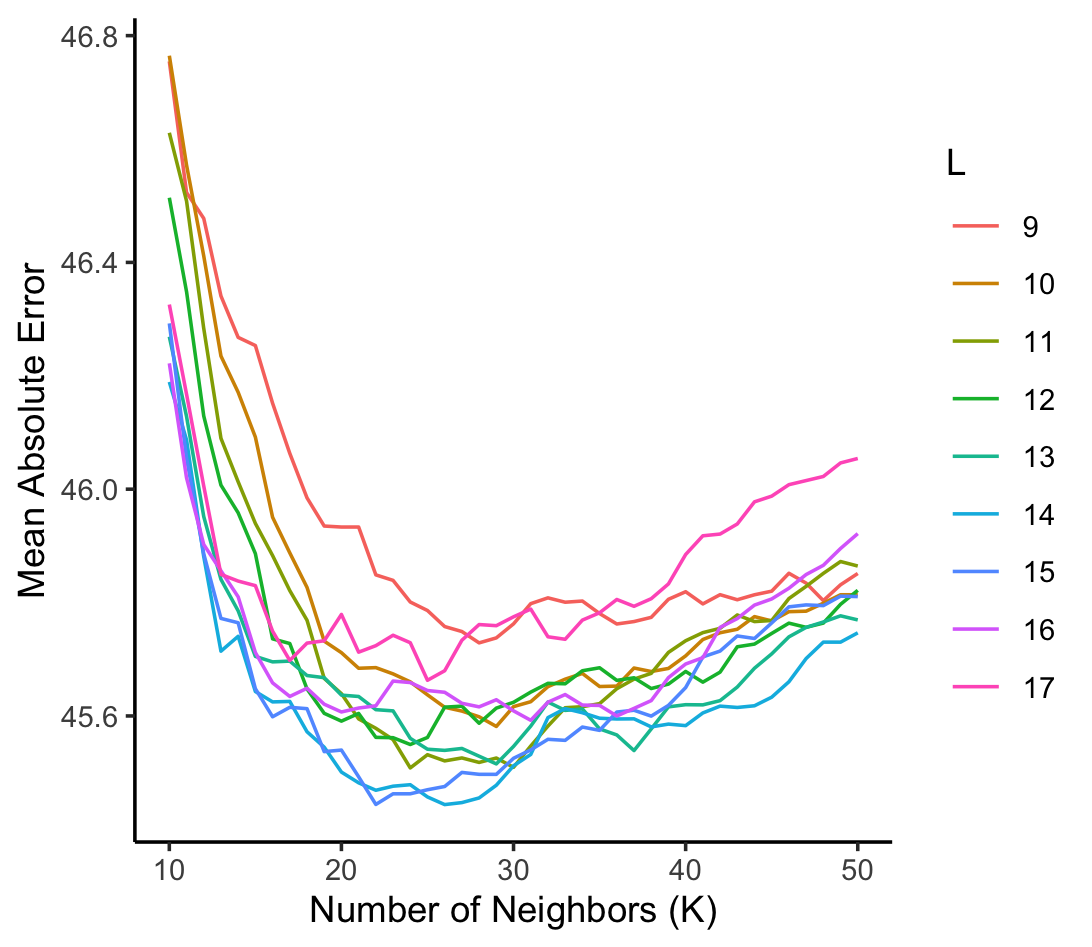}
		\caption{Best $K$ values}
		\label{fig:kl_tick2}
\end{figure}

In Figure \ref{fig:kl_tick1}, we observe that $L = 14$ provides the best results and that there is a steep increase in error after $L > 15$ for all values of $K$. Then, while searching for the optimal $K$ value, a distinguishable tick pattern can be observed in Figure \ref{fig:kl_tick2}. These observations suggest finalizing the hyperparameter search, and set $L = 14$ and $K = 25$ as the optimal values. While analysing the other aspects of the similarity models, we will mainly be using the optimal hyperparameters. However, we will include results with other values of $K$ and $L$ whenever it contributes to the  discussion.

The following table shows the result of the  similarity of trajectories method discussed in this section along with the previously listed benchmark results. Recall that this basic similarity approach was based on  using the weighted Euclidean distance  and  the arithmetic mean of the candidates as forecast.



\begin{table}[H] \centering 
\begin{tabular}{@{\extracolsep{5pt}} ccccccc} 
\\[-1.8ex]\hline 
\hline \\[-1.8ex] 
Model & MAE & MAPE \\ 
\hline \\[-1.8ex] 
Naive & 54.74 / 56.73 & 5.35 / 5.36  \\
AR (9) & 50.34 / 52.62 & 4.92 / 4.94  \\
ARIMA (9,0,3) & 49.34 / 51.83 & 4.82 / 4.83  \\
SARIMA (9,0,3), (3,0,2)   & 42.09 / 46.08 & 4.18 / 4.39  \\
RF & 43.58 / 46.85 & 4.35 / 4.40 \\
Seasonal RF & 42.91 / \textbf{45.99} & 4.31 / \textbf{4.32} \\
Similarity (L = 14, K = 25) & 45.46 / 47.84 & 4.44 / 4.50 \\
\hline \\[-1.8ex] 
\end{tabular} 
  \caption{Preliminary comparison of similarity approach with the best benchmark results}
  \label{tab:point_weuc_res} 
\end{table} 

Table \ref{tab:point_weuc_res} shows that the similarity of trajectories model outperform autoregressive models (without seasonal parameters)  but it performs worse than both Random Forest models and SARIMA. This leads to a motivation for using seasonality in similarity models. 

\vspace{0.1in}

\noindent 
\textbf{Handling Outliers.}
%
%
Extreme values among candidates may lead to biased predictions. Just like any other method,  removing these outliers can improve the performance of forecasting based on similarity of trajectories. In our specific case based on traffic data, it is  observed that the flow values of candidates for certain observations have an interesting distribution. Figure \ref{fig:whale} below shows one such example.

\begin{figure}[H]
	\centering
		\includegraphics[width=.55\textwidth, keepaspectratio]{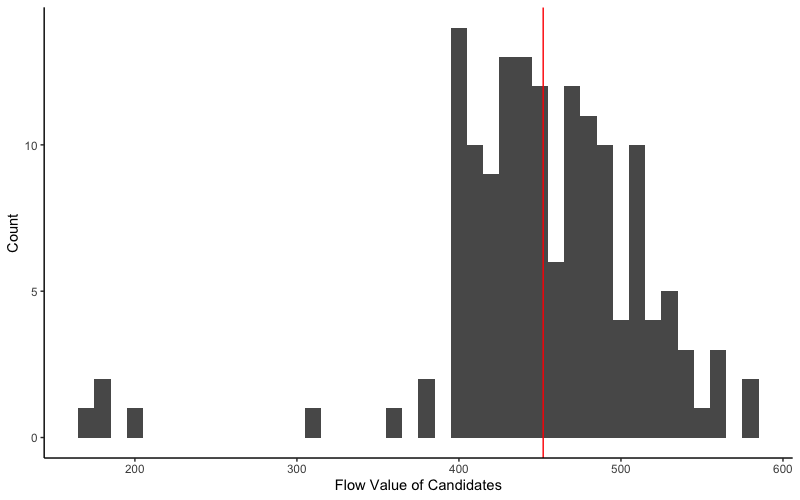}
		\caption{Exemplary distribution of candidate flow values. Red line shows the observed flow.}
		\label{fig:whale}
\end{figure}

In Figure \ref{fig:whale}, observe that there is a cluster of outliers having very low value compared to other candidates. We have observed numerous examples sharing similar candidate distribution and addressing this characteristic may be helpful while handling the outliers. \cite{habt} uses  winsorization for smoothening the extreme values, and they observe that this process  improves their forecasting results. The details of winsorization are discussed in Section \ref{sec:outlier}.  Let us note here that  this  method is   computationally appealing   and computational efficiency should be kept as a priority as this procedure is an extra step.

 The definition of winsorization  deals  only  with the largest and smallest values, but as seen in Figure \ref{fig:whale},   there can be many outliers that need to be taken into account. 
 With this in mind, we experimented with various other approaches towards the outlier values. In particular, it is observed that  using the z-score to capture outliers is slightly more successful than winsorization in our setting. Besides this, a straightforward approach which we call ``Symmetric Tail Removal"   is described in Section \ref{sec:outlier}, again providing small improvements in results. 
 
 Below is a comparison of outlier removal methods with respect to the weighted Euclidean distance, where   optimal hyperparameters are used. Let us note that since some of the outliers are removed,   the optimal value of $K$ may be subject to change. More specifically, here  $L$ is fixed to be 14 and  a   grid search among various values of $K$ is conducted.

\begin{table}[!htbp]
\centering 
\begin{tabular}{@{\extracolsep{5pt}} ccccccc} 
\\[-1.8ex]\hline 
\hline \\[-1.8ex] 
Outlier Method & $L$ & $K$ & MAE \\ 
\hline \\[-1.8ex] 
None & 14 & 25 & 45.46 / 47.84 \\
Winsorization & 14 & 21 & 45.31 / 47.71 \\
Z-Score  & 14 & 27 & 45.21 / 47.54 \\
\textbf{Sym-Tail-Remove} & 14 & 25 & 45.10 / \textbf{47.39} \\
\hline \\[-1.8ex] 
\end{tabular} 
  \caption{Comparison   of outlier methods}
  \label{tab:point_outlier1} 
\end{table} 



The results in Table \ref{tab:point_outlier1} suggest that the use of  simple outlier removal methods may improve the performance slightly. Although we tried several distinct  values for $K$, we have seen that the optimal value tends to be     around 25. We observe that symmetric tail removal is the best method for our case. In the following discussions and the next sections we will only use this method. 

\vspace{0.1in}

\noindent  \textbf{Multi-step forecasting.}  For obtaining multi-step ahead forecasts, a similar strategy to the one-step ahead forecast is used, but  the candidates  are chosen according to the number of steps of interest. Since similarity models are non-parametric and they rely on  use of  patterns, one may hope that  a successful similarity model  perform well even when the step-size is huge.

\begin{table}[!htbp] \centering 
\begin{tabular}{@{\extracolsep{5pt}} ccccccc} 
\\[-1.8ex]\hline 
\hline \\[-1.8ex] 
Model & MAE (Step-2) & MAE (Step-3) & MAE (Step-5) \\ 
\hline \\[-1.8ex] 
Naive & 78.160 / 79.889 & 98.890 / 101.991 & 143.850 / 146.936 \\
AR & 64.325 / 67.552 & 73.414 / 77.103 & 106.138 /111.294 \\
ARIMA & 58.702 / 59.905 & 61.027 / 63.243 & 88.823 / 92.032 \\
RF & 53.975 / 56.480 & 59.056 / 62.658 & 70.866 / 76.639 \\
SRF & 53.571 / 54.836 & 58.260 / 60.568 & 69.863 / 73.694 \\
Similarity & 53.743 / 54.852 & 59.125 / 60.764 & 80.308 / 83.789 \\
Seasonal Similarity & 49.812 / 51.072 & 54.297 / 55.516 & 61.596 / 62.197 \\
\hline \\[-1.8ex] 
\end{tabular} 
  \caption{Multi-step ahead forecasts} \label{tab:step_ahead} 
\end{table}

 Table \ref{tab:step_ahead} demonstrates that the non-seasonal similarity based model performs well when the step size is 2. However, the performance rapidly falls as the step size increases. This could  be  an indication that the model under consideration  misses important patterns from the data. On the other hand, the table contains the results from a ``seasonal similarity" model, details of which to be discussed in Section \ref{sec:filter}, showing Seasonal Similarity model outperforms all of the benchmarks as the step size increases.

\subsection{Prediction Intervals}

Instead of giving a point forecast, providing a prediction interval may be desirable in many scenarios. To the best of our knowledge,   the similarity of trajectories approach has not been   explored  in the literature in this aspect yet. The purpose of this  section is to describe one particular strategy for such a construction.  The idea is simple and can be described  as follows: After obtaining the candidates,  sort them by their value in ascending order and match each observation with the corresponding sample quantile. Afterwards,  the required prediction interval is generated   using the corresponding  quantile forecasts. Note that we use the definition of  the sample quantile given in   Section \ref{subsec:sample_quantile}.

Before proceeding further, let us mention the   benchmark methods that are to be used in the prediction interval framework. A possible natural approach here is to pass from point forecasts to prediction intervals by making use of the sample  standard deviation  of the errors.   
 A second possible option is training the models with quantile loss function. 
There are of course  several  other directions for obtaining prediction intervals, but 
our experiments for benchmarking purposes will be  based on  seven variations around these.

Among these seven,  three of them  use point forecast benchmarks to generate prediction intervals and the other four are Quantile Autoregression (QAR) \cite{koenker_quantile}, Seasonal Quantile Autoregression (SQAR), Quantile Random Forest (QRF) \cite{qrforest} and Seasonal Quantile Random Forest (SQRF) that use quantile loss function. Note that the prediction interval from ARIMA models are only obtained by using sample variance of the point forecast errors.\footnote{Training ARIMA (or moving average models) with quantile loss function does not have a standard definition and we could not find any discussion in the literature. Also, calculation of moving average part using the regression error from a quantile forecast does  not seem to be practical.}

The optimal hyperparameters for quantile autoregressive models are given in Table \ref{table:bench_pi}. QRF and SQRF models are trained simultaneously with their point forecast version using \textit{ranger} R package \cite{ranger_package}. The details regarding hyperparameters can be found on Page \pageref{par:rf_details}, and the R package follows \cite{qrforest} for obtaining quantile predictions. \label{par:rf_pi_details}

We measure performance of these models by using the unconditional coverage and Winkler score defined in Section \ref{sec:evamets}. Table \ref{table:bench_pi} shows that the use of quantile loss function provides significantly better results. In particular, Seasonal Random Forest trained with quantile loss function. Note that (SD) represents the models that are obtained with adding standard deviation to point forecast models and the numbers near QAR and SQAR represent the optimal autoregressive parameters.



\begin{table}[H] \centering 
\begin{tabular}{@{\extracolsep{5pt}} ccccccc} 
\\[-1.8ex]\hline 
\hline \\[-1.8ex] 
 Model & Unconditional Coverage & Winkler Score \\
 \hline \\[-1.8ex] 
 Naive (SD) & 0.9665 / 0.9665 & 426.75 / 420.46 \\
 AR (SD) & 0.9671 / 0.9633 & 391.38 / 389.16 \\
 ARIMA (SD) & 0.9421 / 0.9393 &  378.12 / 373.05 \\
 QAR (9) & 0.9655 / 0.9591 & 368.91 / 369.95 \\
 SQAR (8)(6) & 0.9450 / 0.9386 & 337.50 / 366.81 \\
 QRF & 0.9291 / 0.9548 & 341.08 / 321.89 \\
 SQRF & 0.9349 / 0.9625 & 329.34 / 314.48 \\
\hline \\[-1.8ex] 
\end{tabular} 
\caption{Benchmark results for prediction interval forecasts}
  \label{table:bench_pi} 
\end{table}


%

\noindent
\textbf{Hyperparameter Tuning.}  The optimal value of  the window size $L$  is expected to be similar to the point forecasting case. However, the value of the number of nearest neighbors $K$ may require to be larger in order to capture the underlying distribution better. The initial grid search is done similar to the point forecasting case, with $L$ between 2 and 20, and $K$ between 10 and 200. 

\begin{figure}[H]
	\centering
		\includegraphics[width=.45\textwidth, keepaspectratio]{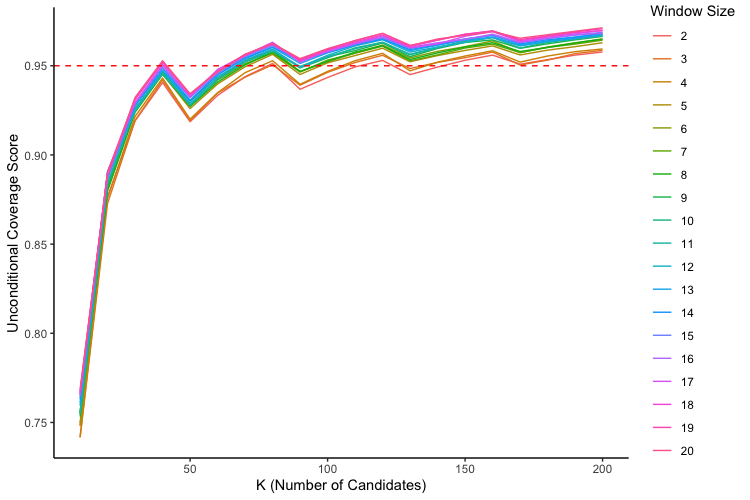}
		\includegraphics[width=.45\textwidth, keepaspectratio]{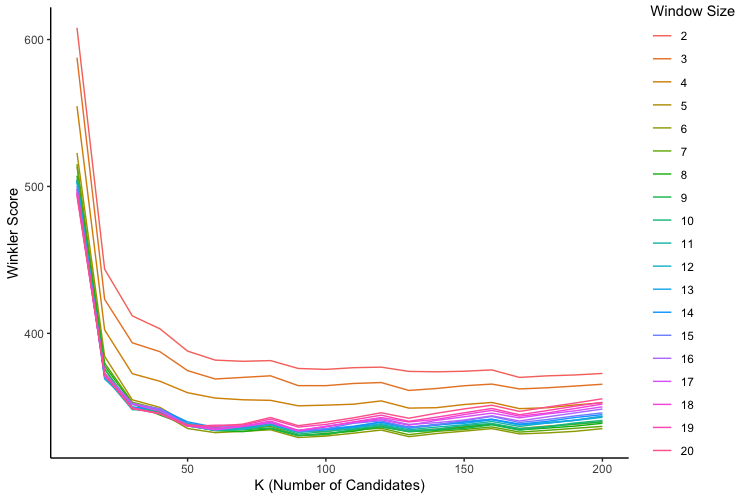}
		\caption{Unconditional coverage and Winkler score}
		\label{fig:uccov_winkler1}
\end{figure}

  Figure \ref{fig:uccov_winkler1} demonstrates that the unconditional coverage stabilizes for $K \geq 70$. Also the Winkler score does not appear to change significantly for $K \geq 50$, for all values of $L$. The best Winkler score is obtained for $L = 9$ and $K = 60$.

\begin{table}[H] \centering 
\begin{tabular}{@{\extracolsep{5pt}} ccccccc} 
\\[-1.8ex]\hline 
\hline \\[-1.8ex] 
 Model & Unconditional Coverage & Winkler Score \\
 \hline \\[-1.8ex] 
  Naive (SD) & 0.9665 / 0.9665 & 426.75 / 420.46 \\
 AR (SD) & 0.9671 / 0.9633 & 391.38 / 389.16 \\
 ARIMA (SD) & 0.9421 / 0.9393 &  378.12 / 373.05 \\
 QAR & 0.9655 / 0.9591 & 368.91 / 369.95 \\
 SQAR  & 0.9450 / 0.9386 & 337.50 / 366.81 \\
 QRF & 0.9291 / 0.9548 & 341.08 / 321.89 \\
 SQRF & 0.9349 / 0.9625 & 329.34 / 314.48 \\
 Similarity & 0.9501 / 0.9451 & 333.85 / 329.50 \\
\hline \\[-1.8ex] 
\end{tabular} 
\caption{Prediction interval  results, similarity approach included.}\label{table:pi_result1} 
\end{table} 

Table \ref{table:pi_result1} shows the comparison between the best similarity model and benchmark models. We see that quantile random forest models are better than the basic  similarity approach of this section. A more detailed treatment  will be given in Section \ref{sec:predictionints}.

\section{Obtaining and Processing Candidates}\label{sec:processing}

\subsection{The design of the query set and reference sets}
\noindent
Let $L\in \mathbb{Z}^+$ be the window size and $T\in \mathbb{Z}^+$ be the size of the traffic flow data $\{x_1,x_2,\ldots,x_T\}$. In this section, by  a \textit{target}, we mean a  \textit{candidate} in our previous setup. Now we design the time series data as trajectories and their targets as follows:

\begin{align*}
    x_1^{\text{traj}}=(x_1,x_2,\ldots,x_L)&,\      y_1^{\text{targ}}=x_{L+1}\\
    x_2^{\text{traj}}=(x_2,x_3,\ldots,x_{L+1})&,\   y_2^{\text{targ}}=x_{L+2}\\
    \vdots\\
    x_{T-L}^{\text{traj}}=(x_{T-L},x_{T-L+1},\ldots,x_{T-1})&,\ y_{T-L}^{\text{targ}}=x_{T}
\end{align*}

We define the query test set as $Q_{\text{test}}=\{[x_{i}^{\text{traj}},y_i^{\text{targ}}]: s\leq i\leq T-L \}$ and the query tune set as $Q_{\text{tune}}=\{[x_{i}^{\text{traj}},y_i^{\text{targ}}]: u\leq i\leq s-1 \}$, where $s$ and $u$ are chosen in our case  to be natural numbers satisfying $T-L-s=s-1-u$ so that we have $|Q_{\text{tune}}|=|Q_{\text{test}}|.$ Note that the data in $Q_{\text{tune}}$ and $Q_{\text{test}}$ correspond to disjoint sets of time intervals. We train models through $Q_{\text{tune}}$ by tuning hyperparameters/parameters and measure accuracies on the test set $Q_{\text{test}}$. Predictions are based on the previously observed trajectories. We pick one trajectory (i.e. $x_{\cdot}^{\text{traj}}$) from $Q_{\text{tune}}$ and define a reference set for it. We also pick one trajectory from $Q_{\text{test}}$ and again define a reference set for it. 
For this purpose, we let $R_{q}^{\text{tune}}=\{[x_i^{\text{traj}},y_{i}^{\text{targ}}]: 1\leq i \leq q-1\}$ a reference set for each $q\in \{u,u+1,\ldots,s-1\}$ and $R_{q^{'}}^{\text{test}}=\{[x_i^{\text{traj}},y_i^{\text{targ}}]: w\leq i\leq q^{'}-1\}$ a reference set for each $q^{'}\in \{s,s+1,\ldots,T-L\}$, where we choose $w$ satisfying $s-1=T-L-w$. Choosing $w$ a particular value subject to this condition guarantees the fact that we consider trajectories with dates going back to at most one year for both the trajectories in the query tune set and the query test set. In fact, putting $w=1$ leads to no problem in practice. However, we preferred to see the results by considering trajectories through a history of equal lengths for both the query tune set and the query test set. Let $q$ be in $\{u,u+1,\ldots, s-1\}$ and $q'$ be in $\{w,w+1,\ldots, T-L\}$ and let $Z_i=[x_{i}^{\text{traj}},y_{i}^{\text{targ}}]$. The following figure demonstrates the construction of these sets. \\
\begin{center}
\includegraphics[scale=0.55]{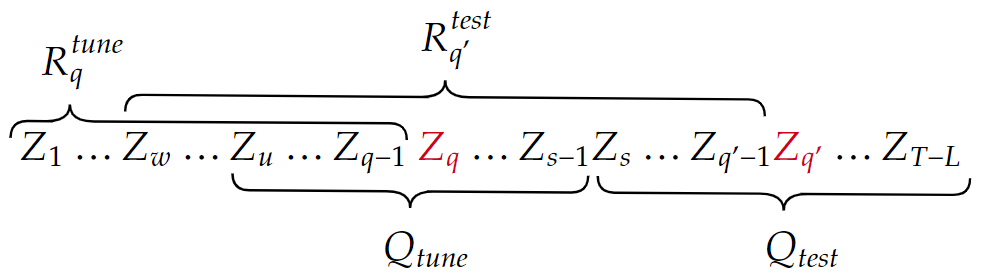}    
\end{center}

\subsection{Obtaining the Nearest Trajectories}\label{sec:obtainnearest}
\noindent
Below we let  $[x_q^{\text{traj}},y_{q}^{\text{targ}}]\in Q_{\text{tune}}$. In general, we are willing to forecast $y_q^{\text{targ}}$ and the prediction provided by the similarity model is denoted by $y_q^{\text{pred}}.$ In this brief subsection we define the necessary notation in order to explain the selection of the nearest trajectories rigorously.

Let $K\in\mathbb{Z}^{+}$ be the number of closest trajectories drawn from $R^{\text{tune,traj}}_q=\{x_{i}^{\text{traj}}: 1\leq i\leq q-1\}$ with respect to some $\textit{distance}$ function $d$. We set $D_q=\{d(x_q^{
\text{traj}
},x):x\in R^{\text{tune,traj}}_q \}$
and denote by $\{d_{q,s}\}_{s=1}^K$ the $K$ smallest numbers of $D_q$ and $\{x^{q,\text{near-traj}}_{s}\}_{s=1}^K \subset R_q^{\text{tune,traj}}$ the $K$ trajectories which are closest to the trajectory $x^{\text{traj}}_q.$ So, $d_{q,s}=d(x_q^{\text{traj}},x^{q,\text{near-traj}}_{s})$ for $s=1,2,\ldots , K.$ Here, without loss of generality we consider $d_{q,s}\leq d_{q,s+1}$ for $1\leq s\leq K-1.$ Finally, that gives us a sub-collection of the reference set $R^{\text{traj}}_q$ defined by $C_q=\{[x^{q,\text{near-traj}}_{s},y^{q,\text{near-targ}}_{s}]:1\leq s\leq K\}$, where $y^{q,\text{near-targ}}_{s}$'s are targets
of $x^{q,\text{near-traj}}_{s}$'s, equipped with the distances $\{d_{q,s}\}_{s=1}^K$ corresponding to $x^{\text{traj}}_q.$

\subsection{Handling Outliers}\label{sec:outlier}

%
%
%
%
%
%
%
%
%
%
%
%
%
%
%
%

Once the nearest neighbors are selected, outlier detection and removal  may as usual help in obtaining better forecasts.  Smoothing the possible outliers via \textit{winsorization} is used in \cite{habt} in setting of similar trajectories.   Below we  briefly discuss the \textit{tail removal} along with winsorization, although we experimented with other approaches as well. 

\vspace{0.1in}

\noindent 
\textbf{Winsorization:} We recall the definition of winsorization  from  \cite{habt}.  Assuming the candidate values for our forecast are given by  $\{f_1, f_2, \dots, f_k \}$ where $f_i \leq f_{i+1}$ for each $i$,  the corresponding winsorized candidates are defined by 
$$
f_{i}^{(w)} = \left\{ \begin{array}{rl}
f_{i+1}, &\mbox{ if $f_i = \min (f_j )$ } \\
f_{i-1}, &\mbox{ if $f_i = \max (f)$ } \\
f_i, & f_i \notin \{\min (f), \max (f)\}
       \end{array} \right.
$$

Simply, this procedure replaces the smallest and highest values of the candidates with the values to them. Another natural approach to outliers is the following.
%

\vspace{0.1in}

\noindent 
\textbf{Tail Removal:} The method removes the smallest $r_1$ and the highest $r_2$ candidates. The values of $r_1$ and $r_2$ can be set in the following ways.

\begin{itemize}
\item (Constant) $r_1  = c_1$ and   $r_2 = c_2$ for some properly chosen  $c_1, c_2$.
\item (Percentile) $r_1 =  \lfloor \gamma_1 K \rfloor$,  $r_2 = \lfloor \gamma_2 K \rfloor$ for  some  positive  $\gamma_1, \gamma_2$ such that $\gamma_1 + \gamma_2 < 1$.
 \end{itemize}
When $\gamma_1 = \gamma_2$, these two can be called as the symmetric constant and symmetric percentile approaches.

\section{Methodologies for point forecasting}\label{sec:pointforecasts}
Recall from Section \ref{sec:obtainnearest} how we obtain $C_q=\{[x^{q,\text{near-traj}}_{s},y^{q,\text{near-targ}}_{s}]:1\leq s\leq K\}$ which is the collection of nearest trajectories and the distances $\{d_{q,s}\}_{s=1}^{K}$ with respect to some distance function $d.$ We use these for point-forecasting in various ways. A basic method for reaching $y^{\text{pred}}_q$ is just a simple averaging:
$$y^{\text{pred}}_q=\frac{1}{K}\sum\limits_{s=1}^Ky^{q,\text{near-targ}}_{s}.$$

One perspective in our approaches below is to provide forecasts using weighted means of targets $y^{q,\text{near-targ}}_{s}$ where the weights are obtained in specific ways that are to be described in Section \ref{sec:wfct}. Another one is making use of local regression which is discussed in  Section \ref{a local regression}. Before proceeding further into these,  let us compare the performance of using  different distance functions and forecasting via just the arithmetic mean of the candidates.\footnote{In Table \ref{tab:comparison_of_metrics} (s) means seasonal filtering and $\mathcal{R}$ represents radius, both of which are to be described in Section \ref{sec:filter}.}

To compare different distance functions, first a grid search for the hyperparameters is conducted on tune set. Even values between 2 and 20 were possible candidates for hyperparameter $L$, and the optimal $K$ value is searched from the set $\{10, 15, 20, \dots, 200\}$. For the seasonal filtered models, $\mathcal{R}$ hyperparameter ranges from 0 to 6.


\begin{table}[H] \centering 
\begin{tabular}{@{\extracolsep{5pt}} cccccccc}
\\[-1.8ex]\hline 
\hline \\[-1.8ex] 
Distance & best hyperp. & MAE & MAPE\\ 
\hline \\[-1.8ex] 
Canberra & $L=8,K=25$ &  46.80 / 48.69 & 4.58 / 4.60 \\
Correlation & $L=20,K=10$ & 146.02 / 130.58 & 13.89 / 11.82 \\
Euclidean & $L=8,K=30$ & 46.40 / 48.40 & 4.54 / 4.57 \\
Head-Tail & $L=8,K=75$ & 48.16 / 49.45 & 4.68 / 4.65 \\
Maximum & $L=8,K=30$ & 47.01 / 49.08 & 4.59 / 4.62\\
W. Euclidean & $L=14,K=25$ & 45.46 / \textbf{47.84} & 4.44 / \textbf{4.50} \\
\hline \\[-1.8ex] 
Canberra (s) & $L=6,K=25,\mathcal{R}=3$ &  45.66 / 46.45 & 4.64 / 4.40 \\
Correlation (s) & $L=20,K=20,\mathcal{R}=2$ & 90.98 / 82.46 & 9.22 / 8.13 \\
Euclidean (s) & $L=8,K=20,\mathcal{R}=3$ & 45.41 / 46.66 & 4.60 / 4.40 \\
Head-Tail (s) & $L=8,K=20,\mathcal{R}=1$ & 46.97 / 47.33 & 4.80 / 4.49 \\
Maximum (s) & $L=8,K=20,\mathcal{R}=5$ & 46.08 / 47.21 & 4.65 / 4.47\\
W. Euclidean (s) & $L=10,K=25,\mathcal{R}=3$ & 44.59 / \textbf{45.82} & 4.51 / \textbf{4.32}\\ 
\hline \\[-1.8ex] 
\end{tabular} 
  \caption{Accuracy comparison with respect to  different distances. The bottom half of the table uses seasonal filtering.}
  \label{tab:comparison_of_metrics} 
\end{table}


Table \ref{tab:comparison_of_metrics} shows that the model using the weighted Euclidean distance is clearly the best among listed. Also, one may observe that applying seasonal filters improves the performance regardless of the distance function used. In the following sections, unless otherwise is stated, the distance function used for obtaining the candidates will be the weighted Euclidean function.



\subsection{Weights for closest trajectories}\label{sec:wfct}

\begin{enumerate}
\item[Model-1] In general, one would expect that if a trajectory from reference set is closer than another one, then its target should be weighted more. Instead of taking the aritmetic mean of corresponding targets we may consider specific weights and take the weighted average:
$$y^{\text{pred}}_q=\sum\limits_{s=1}^Kw_sy^{q,\text{near-targ}}_{s},$$
where $w_s>0$ and $\sum\limits_{s=1}^K w_s=1.$
We define the weights via a non-decreasing function $f:\{1,\ldots, K\}\rightarrow \mathbb{R}$ by setting $w_s=\frac{f(K-s+1)}{S}$, where $S=\sum_{x=1}^{K}f(x).$ For instance, if $f=1$, then the prediction is just the arithmetic mean of the targets. Our experiments involved  the following choices of the functions: $f_1(x)=1,\ f_2(x)=x,\  f_3(x)=\sqrt{x}, \  f_4(x)=\ln (1+x)$ and $f_5(x)=\ln^2 (1+x).$
\item[Model-2] Instead of choosing weights uniformly, we may also customize them according to the distances:  
$$y^{\text{pred}}_q=\sum\limits_{s=1}^Kw_{q,s}y^{q,\text{near-targ}}_{s},$$ where
$w_{q,s}=\frac{g(d_{q,s})}{\sum\limits_{s=1}^Kg(d_{q,s})}$ with $g$  an arbitrary positive decreasing function on $[0,\infty).$ This is based on the intuition that the closer a trajectory is the greater weight of its target should be.
The  functions experimented consisted of the following: $g_1(x)=\frac{1}{x+0.01}, \ g_2(x)=\frac{1}{\sqrt{x}+0.01}, \ g_3(x)=\frac{1}{x\sqrt{x}+0.01}$ and $g_4(x)=\frac{1}{x^2+0.01}$. 
\item[Model-3] We may also determine the weights of $y^{q,\text{near-targ}}_{s}$'s  using linear regression. The linear regression is used  in the following form:
$$y=\sum_{s=1}^{K}w_sy_{s}+w_{K+1}.$$
We find the minimizers as
weights $\tilde{w}_1,\ldots,\tilde{w}_K, \tilde{w}_{K+1}$  and then consider:
$$y^{\text{pred}}_q=\left[\sum\limits_{s=1}^{K}\tilde{w}_s y^{q,\text{near-targ}}_{s}\right ]+\tilde{w}_{K+1}$$
The training is processed using the tune query set. Then, we measure the trained model via test query set.
\end{enumerate}

\subsection{A local regression approach}\label{a local regression}

We may apply linear regression to a local data in contrast to the one discussed in the previous section. Given $x^{\text{traj}}_q,$ we consider $C_q=\{[x^{q,\text{near-traj}}_{s},y^{q,\text{near-targ}}_{s}]:1\leq s\leq K\}$ and set up a linear regression model on $C_q$. If we denote by $x^{q,\text{near-traj}}_{s}(t)$ the $t^{th}$ component of the vector $x^{q,\text{near-traj}}_{s}$ for $s\in\{1,2,\ldots,K\},$ and $x_q^{\text{traj}}(t)$ the $t^{th}$ component of the vector $x_q^{\text{traj}}$, then we indeed minimize the following function
$$L(w_1,w_2,\ldots,w_L,w_{L+1})=\sum\limits_{s=1}^K\bigg|\bigg(\sum\limits_{t=1}^Lw_tx^{q,\text{near-traj}}_{s}(t)\bigg)+w_{L+1}-y^{q,\text{near-targ}}_{s}\bigg|^2,$$
and use the prediction:
$$y^{\text{pred}}_q=\bigg(\sum\limits_{t=1}^L\tilde{w}_tx^{\text{traj}}_q(t)\bigg)+\tilde{w}_{L+1},$$
where $\tilde{w}_1,\tilde{w}_2,\ldots,\tilde{w}_L,\tilde{w}_{L+1}$ are the optimum weights.

\subsection{Variations} \label{sec:variations}

\subsubsection{Filtering \& Seasonality}\label{sec:filter} 

An immediate variation of the basic method can be given by first using a filter on the overall observed trajectories in the corresponding reference set, and then doing a  search for the similar trajectories over the set of remaining candidate trajectories. One particular strategy that can be used is as follows. 
 Start by fixing a trajectory size, say $L_1$, and a distance function $d_1$, and choose   similar trajectories  with respect to $d_1$. Calling the collection of chosen trajectories $R$,   we  may then select the  nearest trajectories from $R$ with respect to another distance function $d_2$ (and with possibly a different window size $L_2$).  

An important related example would be the case where $R$ holds a certain seasonal characteristics. In general, the set $R$ may consist of the trajectories corresponding to previous days, or weeks. This is implicitly done in \cite{habt} where the candidates are only selected from the same time of the  previous days. 

Extending their work, we define a seasonal filter with a radius hyperparameter, which we denote by $\mathcal{R}$. When   $\mathcal{R} = 0$, the set $R$ below will only contain the observations from the same hour and minute of the day. However, as expected, this filter results with  a set $R$ with very few trajectories. This is especially undesirable for making interval forecasts. To increase the number of possible candidates, $\mathcal{R}$ is introduced. This hyperparameter loosens the seasonality condition via appending $R$ by adding the consecutive observations of the observations in $R$. As an example, when $\mathcal{R}$ is 0, for an observation made at 15:00, set $R$ would only contain the observations made at 15:00 on previous days. However, setting $\mathcal{R}$ to 1 would add the observations made at 14:45 and 15:15 to set $R$. Our experiments show that such a modification improves the results significantly, especially in interval forecasting.


\subsubsection{Hourly Model Forecasting}\label{sec:hf}



The underlying patterns in different parts of the day may vary from each other and the optimal hyperparameters may change with respect to these underlying patterns. Thus, using multiple models to make forecasts in different parts of the data could improve overall performance. There are numerous ways to split the data. Time of the day is a simple and reasonable way to split in case of  forecasting traffic flow. Our experiments reveal that such partitioning and using different models for each hour improves the performance significantly. In particular, in our experiments below, for each hour of the day, hyperparameters of the best performing similarity models are selected on tune set.

Let us  provide a brief discussion on the details of experimentation. We used 10-fold cross validation to choose the optimal hyperparameters. For simplicity we fixed the distance to be  the weighted Euclidean, used seasonal filtering, and searched for optimal $L$, $K$ and $\mathcal{R}$ values from 2 to 20, 10 to 200, and 0 to 6, respectively. The grid can be extended to contain other distances. To give a motivation for using multi-hourly models, we provide an hour by hour comparison of some similarity models with different hyperparameters.





\begin{figure}[H]
	\centering
		\includegraphics[width=.95\textwidth, keepaspectratio]{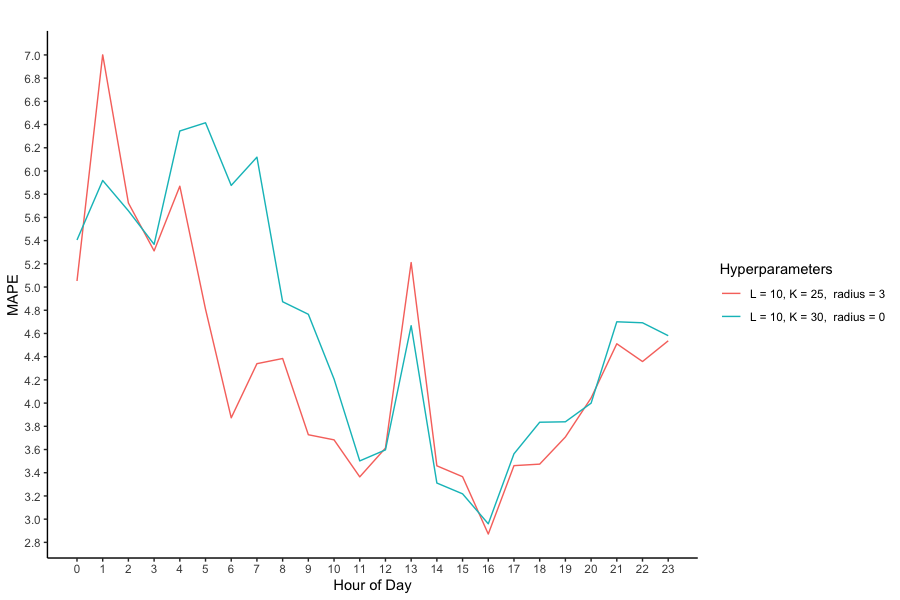}
		\caption{Hourly performance comparison of different similarity models (Point Forecast).}
		\label{fig:hourly_plot2}
\end{figure}

Figure \ref{fig:hourly_plot2} displays an example of performance differences between two different similarity models in various hours of the day. Overall, the model with optimal hyperparameters (shown by red line) performs better. However, in some hours of the day the other model exhibits significantly better performance.

\subsection{Results} \label{subsec:point_results}
Table \ref{fig:point_models} below shows the results of implementations of  Model-1,  Model-2,  Model-3, the Local regression, and their seasonal versions, and lastly the Multi-hourly model, by running a grid search through hyperparameters $L$ and $K$ shown in  Table \ref{fig:point_models_without_seas} below.
The Euclidean (E) and the weighted Euclidean (WE) distances were used for Model-1, Model-2, Model-3, the Local Regression and their seasonal versions Model-1-S, Model-2-S, Model-3-S and Local R.-S. The weighted Euclidean is the only one used for the multi-hourly model.\footnote{In the following tables, $\mathcal{R} = 0$ except the multi-hourly model.}

\begin{table}[H] \centering 
\begin{tabular}{@{\extracolsep{5pt}} cc}
\\[-1.8ex]\hline 
\hline \\[-1.8ex] 
Model & Hyperparameter range\\ 
\hline \\[-1.8ex] 
Model-1 & $L=2,3,\ldots,20$; $K=10,20,\ldots,100$; $f_1,f_2,f_3,f_4,f_5$; E, WE  \\
Model-2 & $L=2,3,\ldots,20$; $K=10,20,\ldots,100$; $g_1,g_2,g_3,g_4$; E, WE \\
Model-3 & $L=2,3,\ldots,19,20$; $K=10,20,\ldots,800$; E, WE\\
Local R. & $L=2,3,\ldots,20$; $K=400,410,\ldots,800$; E, WE\\
Model-1-S & $L=2,3,\ldots,20$; $K=10,20,\ldots,100$; $f_1,f_2,f_3,f_4,f_5$; E, WE\\
Model-2-S & $L=2,3,\ldots,20$; $K=10,20,\ldots,100$; $g_1,g_2,g_3,g_4$; E, WE\\
Model-3-S & $L=2,3,\ldots,20$; $K=10,20,\ldots,360$; E, WE\\
Local R.-S & $L=2,3,\ldots,20$; $K=100,110,\ldots,360$; E, WE\\
Multi-hourly model & $L=2, 3, \dots, 20$, $K =10, 20, \ldots,200$,  $\mathcal{R}= 0, 1, \ldots,6$; WE \\
\hline \\[-1.8ex] 
\end{tabular} 
  \caption{Hyperparameters used for grid search}
  \label{fig:point_models_without_seas} 
\end{table}

\begin{table}[H] \centering 
\begin{tabular}{@{\extracolsep{5pt}} cccc}
\\[-1.8ex]\hline 
\hline \\[-1.8ex] 
Model & best hyperp. & MAE & MAPE\\ 
\hline \\[-1.8ex] 
Naive & - & 54.74 / 56.73 & 5.35 / 5.36\\
AR & $p = 9$ & 50.34 / 52.62 & 4.92 / 4.94 \\
ARIMA  & $p=9,d=0,q=3$ & 49.34 / 51.83 & 4.82 / 4.83\\
SARIMA  & $p=9,d=0,q=3, P=3,D=0,Q=2$ & 42.09 / 46.08 & 4.18 / 4.39\\
RF & See page \pageref{par:rf_details} & 43.58 / 46.85 & 4.35 / 4.40  \\
Seasonal RF & See page \pageref{par:rf_details} & 42.91 / 45.99 & 4.31 / 4.32 \\
Model-1  & $f_2,L=14,K=40$, WE & 45.17 / 47.69  & 4.47 / 4.59 \\
Model-2 & $g_4,L=13,K=40$, WE & 45.08 / 47.59 & 4.39 / 4.47 \\
Model-3 & $L=15, K=780$, WE & 43.82 / 49.88 & 4.33 / 4.70 \\
Local R. & $L=19,K=670$, E & 44.19 / 46.74 & 4.33 / 4.39\\
Model-1-S & $f_2, L=4, K=30$, WE  & 43.23 / 44.39 & 4.43 / 4.23 \\
Model-2-S & $g_3, L=4, K=40$, WE & 43.21 / 44.36 & 4.43 / 4.23 \\
Model-3-S & $L=4, K=350$, WE & 43.01 / 45.39 & 4.40 / 4.32 \\
Local R.-S & $L=5,K=260$, E & 41.13 / \textbf{43.16} & 4.08 / \textbf{4.09}\\
Multi-hourly model & Multiple Models & 42.88 / \textbf{43.28} & 4.35 / \textbf{4.09} \\
\hline \\[-1.8ex] 
\end{tabular} 
  \caption{Best results  of the models for  certain choices of hyperparameters.}
  \label{fig:point_models} 
\end{table}
 


It is seen from Table \ref{fig:point_models} that the local regression (with similarity) and the multi-hourly model performs better than both the benchmark and other similarity based methods. In particular, let us note that  the similarity approach of Model-1-S was used  in \cite{habt}. 

\begin{figure}[H]
	\centering
		\includegraphics[width=.6\textwidth, keepaspectratio]{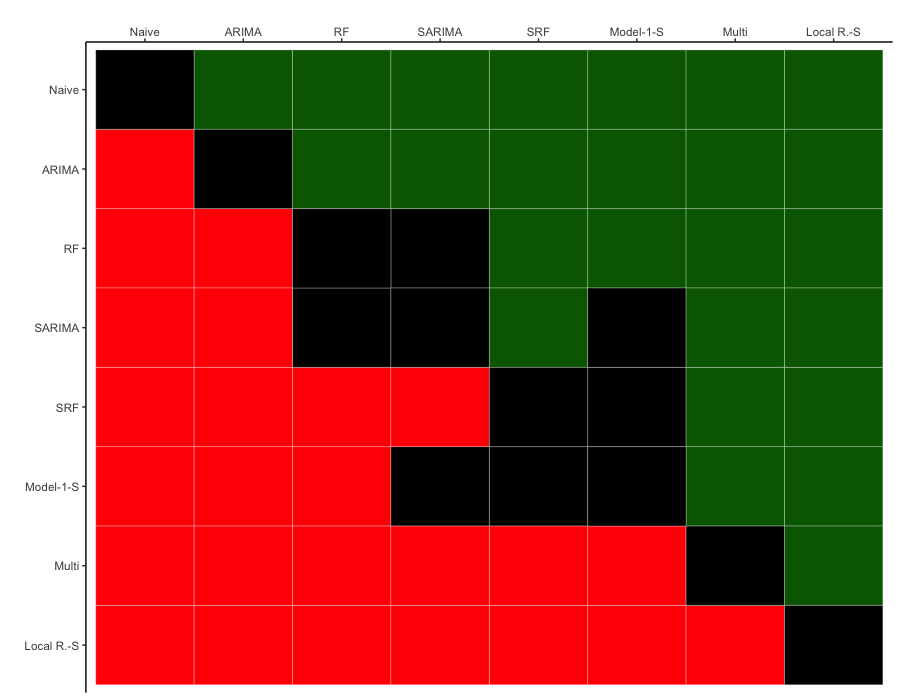}
		\caption{Diebold-Mariano test results. Dark green means column-model is better than row-model. Black means no significant superiority can be deduced.}
		\label{fig:dm_test1}
\end{figure}

Also, one may also like to see whether this conclusion can be shown to hold via a statistical test. When we used the Diablod-Mariano test to compare the benchmarks, we have observed that seasonal random forest model is significantly the best benchmark model. Comparing Model-1-S with benchmarks revealed that Model-1-S is significantly better than the naive, ARIMA and random forest models. However, we have observed no superiority comparing Model-1-S with SARIMA and seasonal random forest. Multi-hourly model significantly outperforms the benchmarks and Model-1-S. Finally, seasonal local regression model exhibited a superior performance against all of the other models shown in  Table \ref{fig:point_models}.





\section{Prediction intervals based on similarity}\label{sec:predictionints}

The purpose of this section is to  discuss our  approaches on obtaining  prediction intervals  using similarity of trajectories. The section begins with a general discussion on existing methods in the literature, and then introduces  two approaches which we  call ST (Similarity Trajectory) and MDST (Model Dependent Similarity Trajectory).


\subsection{Background}

Let $X_t$, $t \geq 1$, be  the time series of interest.  
 One naive standard strategy for obtaining a prediction interval is the historical simulation  approach. In this case, there is some underlying model (e.g., ARIMA, LSTM,...) that provides point forecasts for the next time instance. Let us write $F_t$ for the model predictions provided by this model. Now suppose that we are at some time $T \in \mathbb{N}$, and that we would like to obtain a prediction interval for time  $T+1$ based on observations corresponding to times $1,\ldots,T$.\footnote{As in point forecasts, our prediction intervals will be constructed  solely based on the earlier observations for the time series of interest, and there will be no exogenous data used. Variations that use similarity of certain other data could improve the results here. In case of the traffic forecasting problem, for example, the data obtained from some other station could be considered as  the exogeneous data.} For some predetermined $L \in \mathbb{N}$, the historical simulation approach begins by considering the actual values $X_t$, and the model forecasts $F_t$, for $t \in \{T - L + 1,\ldots,T\}$, and looks at the corresponding errors $\epsilon_t = F_t - X_t$. Then, for given $\alpha \in (0,1)$, in order to obtain a $(1- \alpha ) 100 \%$ prediction interval  for $X_{T + 1}$, we choose the sample quantiles $\epsilon(\alpha / 2)$, $\epsilon(1 - \alpha / 2)$ of the sequence $\epsilon_{T - L + 1},\ldots, \epsilon_{T }$. Then the $(1- \alpha ) 100 \%$ prediction interval is given by $$(F_{T+1} + \epsilon(\alpha / 2), F_{T+1} + \epsilon( 1- \alpha / 2)).$$
We may summarize the discussion on historical simulation prediction intervals, which we call HS, as follows:
\begin{itemize}
\item[HS1.] Consider the errors $\epsilon$ made done once a specific model is used for point forecasts.   
\item[HS2.] Fix some window size $K$, and consider the most recent $K$ errors. Choose the corresponding  $\alpha / 2$ and  $ 1- \alpha / 2$ sample quantiles. 
\item[HS3.] Construct the  prediction interval by adding  these quantiles to the point forecasts provided by the model. 
\end{itemize}
The choice of $L$ in this standard setup can  be done  by using grid search. Also let us note that a more reasonable way to choose the errors to compute the quantiles could be the use of seasonality. For example, instead of looking at the previous $L$ errors, one could go back further, and look at the $L$ errors corresponding to that instance of previous $L$ days. Such an approach will be called the seasonal historical simulation (HS-S).


Below, we will be using the HS method as a benchmark method in our comparisons. However, there are  various other techniques for obtaining prediction intervals that are not discussed below. These include but are not restricted to the distributional prediction interval approach, bootstrap method,  and the quantile regression averaging (QRA). Among these, QRA, in which the regressors are the forecasts of certain individual models, is of particular interest to us. In an independent work \cite{adi:2023} in preparation, we combine the basic ideas behind QRA and the similar trajectories, and introduce some novel  methodologies for providing prediction intervals.

\subsection{Similarity based prediction intervals}

In this subsection we propose two methods for obtaining prediction intervals. The following flowchart provides a general guideline for the tools that are used in these two approaches: ST and MDST. 


\begin{figure}[H]
	\centering
		\includegraphics[width=.9\textwidth, keepaspectratio]{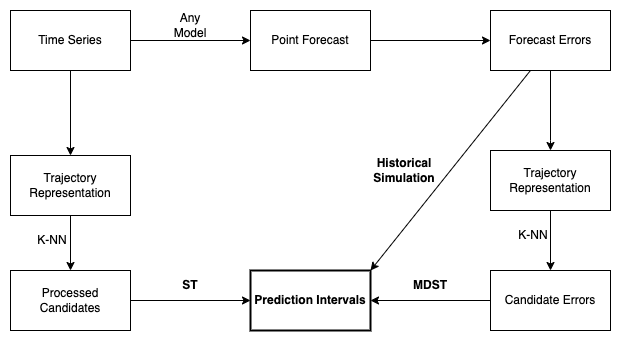}
		\caption{Summary of ST and MDST}
		\label{fig:flowchart2}
\end{figure}

The first method we propose  is analogous to the case of similarity based point forecasting, and it will be called the similarity of trajectories (ST) approach. It consists of the following steps: 
\begin{itemize}
\item[ST1.] Fix a distance function, a window size, and choose $K$ most similar trajectories in the past.
\item[ST2.] Consider the set of possible candidates $\{f_1,\ldots,f_K\}$, corresponding to these $K$ similar trajectories,  as the actual forecast.   
\item[ST3.] Compute the $f(\alpha / 2)$, $f(1 - \alpha / 2)$ sample quantiles and form the prediction interval as $$(f(\alpha / 2), f(1 - \alpha / 2)).$$
\end{itemize}
Note that the described method is different than the historical simulation approach in the sense that it is independent of any model assumptions. No probabilistic assumptions are made, and the given methodology is totally deterministic in nature. When the trajectories are selected among a filtered data according to seasonality (Section \ref{sec:filter}), the method will be called ST-S, and when the forecasting is done on an  hourly basis as in Section \ref{sec:hf}, it will be called multi-hourly model.

Beyond the ST approach,  one may wonder how we  may combine the HS and ST methods for obtaining prediction intervals. Towards this direction, we introduce and experiment on the following strategy, which we call the model dependent similarity trajectory (MDST):  
\begin{itemize}
\item[MDST1.]  Consider the error sequence $\epsilon$ discussed in HS approach. 
\item[MDST2.]   Fix a distance function, a window size, and choose $K$ most similar trajectories in the past, with respect to the corresponding part of the error sequence.
\item[MDST3.]  Consider the set of possible candidates $\{\epsilon_1,\ldots,\epsilon_K\}$ for the errors. 
\item[MDST4.] Form the prediction interval by adding the  $\alpha / 2$ and  $ 1- \alpha / 2$ sample quantiles of the candidate errors to the point forecast provided by the model.
\end{itemize}

When daily seasonal seasonality is involved, the method  will be called MDST-S. This seasonal variation  includes the trajectories only starting from the same time of the day as possible candidates,  as described in Section \ref{sec:filter}. This approach improves the performance significantly as can be seen in the Table \ref{table:pi_result_final}.

\subsection{Results}
  
The purpose of this section is to experiment on the performance of  using similarity of trajectories in obtaining interval forecasts. Table \ref{table:pi_result_final} compares the benchmark and similarity based models using unconditional coverage and the Winkler score. 
The models are trained to make interval forecasts with $ 95 \% $ coverage.
While unconditional coverage shows the accuracy of the models in terms of covering $ 95 \%$ of the observations, the Winkler score is helpful for comparing these models with each other.

\begin{table}[H] \centering 
\begin{tabular}{@{\extracolsep{5pt}} ccccccc} 
\\[-1.8ex]\hline 
\hline \\[-1.8ex] 
 Model & Best Hyperp. & Unconditional Coverage & Winkler Score \\
 \hline \\[-1.8ex] 
 Naive & --- & 0.9665 / 0.9665 & 426.75 / 420.46 \\
 QAR & $p = 9$ & 0.9655 / 0.9591 & 368.91 / 369.95 \\
 SQAR & $p = 8, P = 6$  & 0.9450 / 0.9386 & 337.50 / 366.81 \\
 QRF & See page \pageref{par:rf_pi_details} & 0.9443 / 0.9548 & 315.75 / 321.89 \\
 SQRF & See page \pageref{par:rf_pi_details} &  0.9433 / 0.9625 & 304.95 / 314.48 \\
 HS & $L = 60$ & 0.9145 / 0.9168 & 388.46 / 374.24 \\
 HS-S & $L = 60$ &  0.9165 / 0.9208 & 335.26 / 327.11 \\
 ST & $L = 9, K = 60$  & 0.9427 / 0.9358 & 333.85 / 329.50 \\
 ST-S  & $L = 4, K = 150, \mathcal{R} = 5$  & 0.9499 / 0.9457 & 349.43 / 317.32 \\
 Multi-hourly Model & Multiple Models & 0.9498 / 0.9424 & 335.99 / 298.44 \\
 MDST & $L = 8, K = 220$ & 0.9301 / 0.9308 & 340.35 / 326.20 \\
 MDST-S & $L = 8, K = 220, \mathcal{R} = 6$ & 0.9410 / 0.9406 & 320.53	 / 314.56 \\
\hline \\[-1.8ex] 
\end{tabular} 
\caption{Prediction interval performance comparison}
  \label{table:pi_result_final} 
\end{table} 

The results from Table \ref{table:pi_result_final} suggests that  the best approach among listed is the  multi-hourly model which is the ST-S method in which forecast models are obtained in an hourly basis. None of the other similarity models were better than seasonal random forest but we have seen that the MDST-S and seasonal random forest model exhibit competitive performance with each other.

\begin{figure}[H]
\centering	\includegraphics[width=.6\textwidth, keepaspectratio]{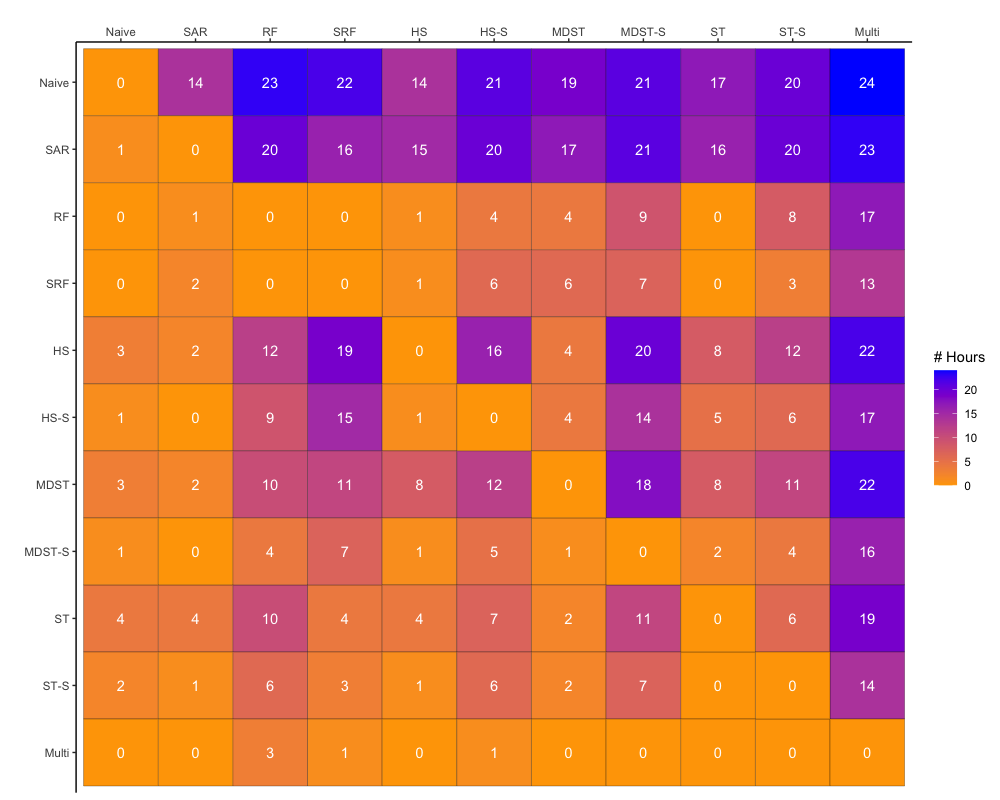}
\caption{Diebold Mariano Test - Hour based comparison for prediction intervals.}
		\label{fig:dm_test2}
\end{figure}


Also, one may also like to see whether these conclusions can be shown to hold via a statistical test. Following \cite{NW:2018}, we used Diebold-Mariano test for comparing the performance of interval forecasts via their Winkler score. First, we conducted the test on the whole data (as in Section \ref{subsec:point_results}). The test resulted in multi-hourly model significantly outperforming all other models and among the benchmarks, seasonal random forest was the best model. Also, MDST-S turns out to  outperform all models except multi-hourly model and seasonal random forest, and no conclusions could be drawn  between MDST-S and seasonal random forest.

Secondly, using the hour based Diebold-Mariano test approach, again similar to \cite{NW:2018}, we have seen that multi-hourly model performs better than all other models in most hours of the day. As demonstrated in Figure \ref{fig:dm_test2}, following the multi-hourly model, MDST-S, ST-S and seasonal random forest were the successful ones. Lastly, let us note that 
multi-hourly model was superior against seasonal random forest in only 13 hours of the day, and it was significantly worse in 1 hour  of the day.  So the test resulted inconclusive for 10 hours. The competition between the two models in these hours of the day motivates the search for possible conditions where similarity models work well and the ones open to improvement. 

%
%
%
%
%
%
%
%
%
%
%
%
%
%

\section{Conclusion}\label{sec:conclusion}

The literature on time series forecasting, and in particular  on traffic flow forecasting, is very rich, with various approaches and methodologies. Our motivation in this   work was to have a general look at  the   use of  similarity of trajectories, and to discuss its possible variations. The experiments consisted of obtaining prediction intervals together with point forecasts. 

Our conclusions above can be summarized as follows: 
\begin{itemize}
\item The methods used in this manuscript  seem to yield competitive results against the more sophisticated  models such as the random forest. The similarity of trajectories approach  has a great flexibility and provides the chance to explain the forecast by a direct reference to historic behavior.   
\item Using local regression and including seasonality improve the point forecast results significantly. However, each additional  step comes up with new hyperparameters and computational issues, and these can be analyzed in an upcoming work.
\item Multi-step forecasts using seasonal similarity seem to perform very well, and this can be further analyzed in a separate experimental  study. 
\item Obtaining prediction intervals based on similar trajectories in the same manner seem to be a natural choice, and it yielded results comparable to the ones we used as benchmarks,  but we believe that the  performance of the proposed strategies are  open to improvements. 
\item
Furthermore, in point forecasting, we may customize local regression  by replacing linear regressing with other machine learning algorithms (e.g., gradient boosting, support vector machhine) for locally obtained data.  
\end{itemize}

In conclusion, the methods we surveyed around are flexible, and are open to further improvements. Our plan in subsequent work is to specialize on certain aspects of the problem and develop the current strategies in different perspectives via the use of  techniques from the statistics and machine learning literature.

 \bigskip 
 
 \textbf{Acknowledgements} The authors would like to thank Elif Yılmaz for helpful discussions. Second and third authors are supported partially by BAP grant 20B06P.

\end{document}